\documentclass[sigconf]{acmart}

\AtBeginDocument{%
  }
\AtBeginDocument{%
  }
\usepackage{booktabs}
\usepackage{multirow}
\usepackage{arydshln}  
\usepackage[normalem]{ulem}
\usepackage{graphicx}
\usepackage{subcaption}  
\useunder{\uline}{\ul}{}
\usepackage{listings}
\usepackage{fontawesome}
\usepackage{multirow}
\usepackage{arydshln}
\usepackage{comment}
\usepackage{subcaption}
\usepackage{booktabs}
\usepackage{cancel}
\usepackage{caption}
\usepackage{titlesec}
\usepackage{balance}

\titlespacing*{\section}{0pt}{1.5ex plus .0ex minus .0ex}{1.5ex plus .0ex}
\titlespacing*{\subsection}{0pt}{1.5ex plus .0ex minus .0ex}{1.5ex plus .0ex}

\copyrightyear{2025}
\acmYear{2025}
\setcopyright{acmlicensed}\acmConference[MM '25]{Proceedings of the 33rd ACM International Conference on Multimedia}{October 27--31, 2025}{Dublin, Ireland}
\acmBooktitle{Proceedings of the 33rd ACM International Conference on Multimedia (MM '25), October 27--31, 2025, Dublin, Ireland}
\acmDOI{10.1145/3746027.3755777}
\acmISBN{979-8-4007-2035-2/2025/10}

\begin{document}

\title{\textbf{EEmo-Bench}: A Benchmark for Multi-modal Large Language Models on Image Evoked Emotion Assessment}

\author{Lancheng Gao}
\affiliation{%
  \institution{Shanghai Jiao Tong University}
  \city{Shanghai}
  \country{China}
}
\email{gaolancheng@sjtu.edu.cn}

\author{Ziheng Jia}
\affiliation{%
  \institution{Shanghai Jiao Tong University}
  \city{Shanghai}
  \country{China}}
\email{jzhws1@sjtu.edu.cn}

\author{Yunhao Zeng}
\affiliation{%
  \institution{Shanghai Jiao Tong University}
  \city{Shanghai}
  \country{China}
}
\email{SJTU.Zzyh@sjtu.edu.cn}

\author{Wei Sun}
\affiliation{%
 \institution{East China Normal University}
 \city{Shanghai}
 \country{China}}
\email{wsun@cee.ecnu.edu.cn}

\author{Yiming Zhang}
\affiliation{%
 \institution{Shanghai Jiao Tong University}
 \city{Shanghai}
 \country{China}}
\email{ming\_zhang\_sjtu@sjtu.edu.cn}

\author{Wei Zhou}
\affiliation{%
  \institution{Cardiff University}
  \city{Cardiff}
  \country{UK}}
\email{zhouw26@cardiff.ac.uk}

\author{Guangtao Zhai}
\affiliation{%
 \institution{Shanghai Jiao Tong University}
 \city{Shanghai}
 \country{China}}
\email{zhaiguangtao@sjtu.edu.cn}

\author{Xiongkuo Min$\dag$}
\authornote{Corresponding authors$\dag$.}
\affiliation{%
 \institution{Shanghai Jiao Tong University}
 \city{Shanghai}
 \country{China}}
\email{minxiongkuo@sjtu.edu.cn}

\renewcommand{\shortauthors}{Lancheng Gao et al.}

\begin{abstract}
  The furnishing of multi-modal large language models (MLLMs) has led to the emergence of numerous benchmark studies, particularly those evaluating their perception and understanding capabilities. 
 Among these, understanding image-evoked emotions aims to enhance MLLMs' empathy, with significant applications such as human-machine interaction and advertising recommendations. However, current evaluations of this MLLM capability remain coarse-grained, and a systematic and comprehensive assessment is still lacking.
  To this end, we introduce \textbf{EEmo-Bench}, a novel benchmark dedicated to the analysis of the \underline{e}voked \underline{emo}tions in images across diverse content categories.
  Our core contributions include:
  1) Regarding the diversity of the evoked emotions, we adopt an emotion ranking strategy and employ the Valence-Arousal-Dominance (VAD) as emotional attributes for emotional assessment. In line with this methodology, $1,960$ images are collected and manually annotated.
  2) We design four tasks to evaluate MLLMs' ability to capture the evoked emotions by single images and their associated attributes: \textbf{Perception}, \textbf{Ranking}, \textbf{Description}, and \textbf{Assessment}. Additionally, image-pairwise analysis is introduced to investigate the model's proficiency in performing joint and comparative analysis.
  In total, we collect $6,773$ question-answer pairs and perform a thorough assessment on $19$ commonly-used MLLMs.
  The results indicate that while some proprietary and large-scale open-source MLLMs achieve promising overall performance, the analytical capabilities in certain evaluation dimensions remain suboptimal.
  Our \textbf{EEmo-Bench} paves the path for further research aimed at enhancing the comprehensive perceiving and understanding capabilities of MLLMs concerning image-evoked emotions, which is crucial for machine-centric emotion perception and understanding. 
  Our code and benchmark datasets are available at https://github.com/workerred/EEmo-Bench.

\end{abstract}



\maketitle

\begin{figure}
  \centering
  \includegraphics[width=0.9\linewidth]{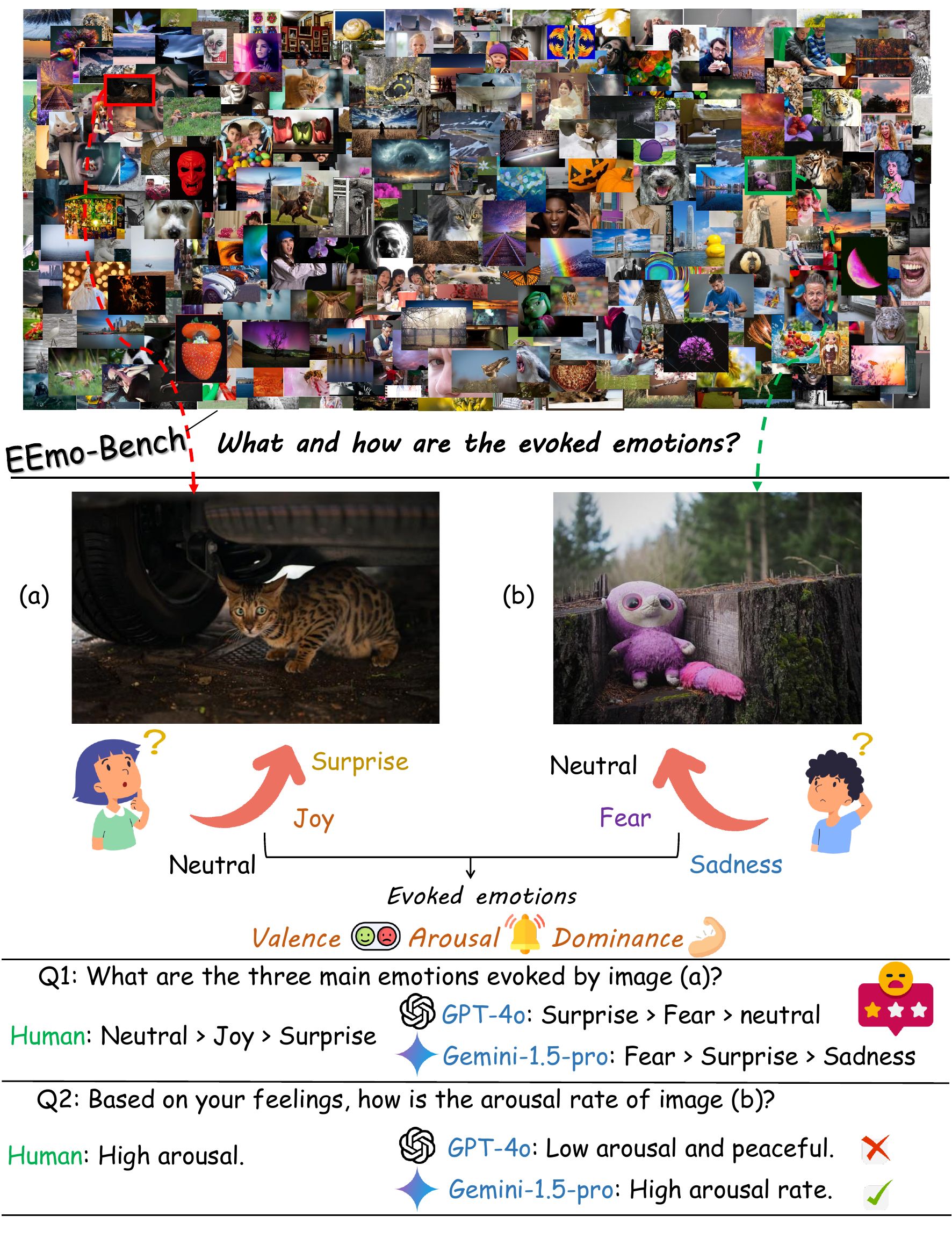}
  \vspace{-8pt}
  \caption{Illustration of the research topic and motivations. \textbf{EEmo-Bench} is focused on the diversity of evoked emotions and the associated attributes, involving valence, arousal, and dominance, providing a comprehensive emotional analysis.}
  \Description{Spot-light}
  \label{fig:spot-light}
\end{figure}

\section{Introduction}


Understanding image-evoked emotions is crucial for extracting the implicit content from images, especially those without explicit expressive themes. 
\textbf{Assessing this capability has wide-ranging applications, particularly in enhancing MLLMs' empathy and emotional resonance} \cite{human-robot-interaction}, which can benefit areas such as human-machine interaction \cite{human-robot}, advertising recommendations \cite{advertising}, public opinion monitoring \cite{social_discuss}, and image generation \cite{generation}.
However, the analysis of image-evoked emotions still faces significant challenges, primarily in two aspects:
1) Viewer responses can vary significantly due to individual divergence, and even a single viewer may experience multiple emotional reactions, emphasizing the need for nuanced and diverse emotional representation. \cite{Emotion6, AC}.
2) Evoked emotions often overlap with the emotions expressed by the image's subject, a phenomenon that is derived from empathy and emotional resonance \cite{zhou2003empathy, Empathy_facial}, leading to confusion for viewers in determining their primary emotional response \cite{EmoSet}.

Recent advancements in multi-modal large language models (MLLMs), such as GPT-4o \cite{GPT4} and Gemini-1.5-pro \cite{gemini}, have demonstrated exceptional performance in perception and understanding tasks across various benchmarks \cite{Survey}.
However, Image Emotion Analysis (IEA), often being explored in the Image Aesthetic Assessment (IAA) \cite{Aesbench, UNIAA}, still  has drawbacks to be completely solved. Specifically, they remain largely coarse-grained and lack comprehensive evaluation: 
1) Many datasets employ emotion classification methods that are neither widely applicable nor grounded in theoretical frameworks, using terms that are rarely recommended for professional use, such as `vitality' and `mystery'. Additionally, most studies denote overall emotional responses with a dominant emotion, which is insufficient for comprehensive emotional representation.
2) Existing benchmarks primarily focus on valence, classifying emotions in a coarse-grained  manner as positive or negative while neglecting essential dimensions like arousal and dominance \cite{IAPS}, leading to an incomplete and less systematic understanding of emotional properties.

To bridge these gaps, we introduce \textbf{EEmo-Bench}, a novel benchmark dedicated to systematically evaluating the abilities of MLLMs in perceiving and understanding \underline{e}voked \underline{emo}tions from images.
Through a systematic study of the theoretical frameworks in the IEA field, EEmo-Bench employs seven fundamental emotions based on Ekman's basic emotions \cite{ekman1982emotion} and incorporates the Valence-Arousal-Dominance (VAD) model \cite{IAPS, VAD_norm} as core emotional attributes, which are widely used in traditional emotion analysis. 
In terms of the diversity of emotional responses, we apply a ranking
strategy for explicit evaluation. The three main evoked emotions for each image are retained and ranked in descending order based on intensity and salience.  Unlike emotion distribution methods \cite{Emotion6}, this ranking strategy is better suited for MLLMs, which often struggle to generate balanced distributions \cite{not_distribution}.
In total, we create a dataset of $1,960$ images with broad content categories, which are manually annotated and strictly scrutinized with the aforementioned emotional attributes.


Furthermore, a four-task evaluation framework is established to enable a comprehensive and fine-grained assessment:
a) \textbf{Perception}. This task employs different types of questions to evaluate MLLMs’ perception ability of evoked emotions in both single and paired image analyses.
b) \textbf{Ranking}. It is designed to measure MLLMs’ proficiency in identifying and sorting different evoked emotions by intensity evoked by an image, utilizing a novel estimation methodology.
c) \textbf{Description}. This task assesses MLLMs’ descriptive  ability to generate detailed emotional descriptions and conduct attributive analysis through open-ended questions, emphasizing chain-of-thought (CoT) reasoning capabilities.
d) \textbf{Assessment}. This task evaluates MLLMs’ capability to quantitatively predict VAD attributes using an adjective-based rating scheme.

Through comprehensive experiments, we observe that while MLLMs demonstrate promising performance on fundamental course-grained tasks, they still face significant limitations in completely and comprehensively understanding the evoked emotions.
The contributions of this work can be summarized as three-fold. 

\begin{itemize}
    \item We introduce \textbf{EEmo-Bench}, the first comprehensive benchmark designed to evaluate MLLMs' ability to perceive and analyze image-evoked emotions. This benchmark includes a meticulously curated dataset of $1,960$ images and $6,773$ manually crafted question-answer (Q\&A) pairs.
    \item We establish a comprehensive evaluation framework comprising four tasks: perception, ranking, description, and assessment, providing a thorough analysis of the adopted emotional attributes. Additionally, pairwise analysis is also conducted to evaluate MLLMs' evoked emotion understanding capabilities with joint or comparative image analysis.
    \item \textbf{EEmo-Bench} incorporates a comprehensive evaluation of $19$ prominent open-source or proprietary MLLMs. The results demonstrate that while \textbf{Qwen2.5-VL-72B} achieves the best overall performance, there remains a substantial gap in comprehensive and fine-grained emotion understanding, providing compelling insights for future advancements.
\end{itemize}


\begin{figure*}
  \includegraphics[width=0.98\textwidth]{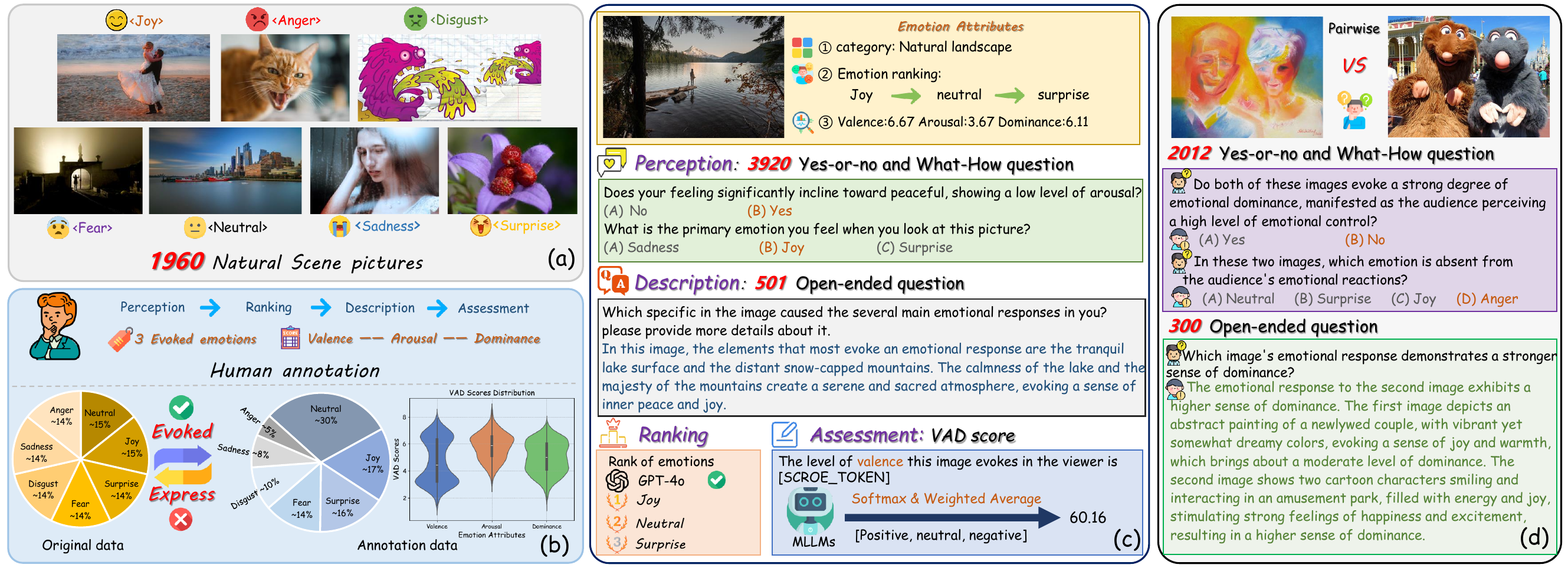}
  \vspace{-8pt}
  \caption{The construction pipeline of the proposed \textbf{EEmo-Bench}.
  To ensure diversity across content categories, we collect images from \textit{Flickr} based on emotional keywords as shown in (a) and manually re-annotate them with emotional properties, including evoked emotion ranking, valence, arousal, and dominance as exhibited in (b).
  As illustrated in (c) and (d), we further conduct four-dimensional tasks for single images and two-dimensional tasks for image pairs, enhancing the systematicity and comprehensiveness of the evaluation process.
  }
  \Description{pipeline}
  \label{fig:pipeline}
    \vspace{-0.3cm}
\end{figure*}

\section{Related Work}

\subsection{Attributes for Image Emotion analysis}

With the growing availability of IEA datasets, the emotion categories involved in them have become more varied. Some datasets use coarse-grained classifications of $6$ to $8$ emotions \cite{Artphoto, Emotion6, 8-emotion, ArtEmis}, based on traditional theories such as Ekman’s basic emotions \cite{ekman1982emotion}, while others categorize emotions into $26$ or $27$ types for more detailed analysis \cite{26_emotions, 27_emotions, self-27}. 
In terms of application, the former provides clearer and more representative categorization, making it ideal for the emotion  classification of diverse images, while the latter is more meticulous and precise, especially utilized in fields such as aesthetic contexts \cite{Aesbench, UNIAA, wikiart}.
Despite the clear definition of emotion categories, individual biases in preferences, experiences, and backgrounds can lead to diverse emotional reactions \cite{diversity}, complicating the understanding of image-evoked emotion denotation.
In response to this, most studies utilize the dominant emotion in representing the emotion response \cite{FI-1, FI-2, interpretable, Internet}, while some attempt to capture a broader spectrum through emotion distributions \cite{Emotion6} or descriptive captions \cite{Affection}, providing a more comprehensive emotional representation.
In addition to categorization, emotional attributes such as valence \cite{FI-1, FI-2}, arousal \cite{VA, AffectNet}, dominance \cite{IAPS, VAD_norm}, and normative significance \cite{GAPED} have been widely studied, offering a deeper perspective on emotions. 
Despite the above efforts, there is still a notable gap in existing theories and datasets on applying widely used classical emotional attributes and extending analytical tasks, while both of them are essential for a comprehensive evaluation of emotions.


\subsection{Benchmarks for Image Emotion Analysis}

The rapid progress of MLLMs has boosted the development of benchmarks evaluating their perception and understanding capabilities across various visual quality assessment domains, including image quality assessment (IQA) and IAA \cite{Survey, MME, MME-SAP, Q-bench, Q-BENCH*, Q-instruct, Q-align, Aesbench, UNIAA}.
Emotion assessment, as a crucial component for evaluating MLLMs' high-level understanding abilities, is first introduced in Large Language Model (LLM) benchmarks \cite{text-1, text-2}, with numerous works emerging, such as EQ-Bench \cite{eqbench} and EmotionQueen \cite{EmotionQueen}. 
However, text-based emotion analysis relies on perceiving emotions through keywords or descriptions, which fundamentally differs from IEA, which involves more complex perceptual mechanisms.
Thus, a model's text-based emotional understanding ability does not accurately reflect its performance in the IEA evaluation.
In response to this, early IEA works in IAA apply Q\&A tasks designed to probe the dominant emotion and its causes in images, assessing MLLMs' foundational ability to understand image emotions \cite{Aesbench, UNIAA, Aesexpert}. However, there is still a lack of fine-grained and specified image-evoked emotion evaluation benchmarks to thoroughly evaluate MLLMs' capability to fully perceive and analyze image-evoked emotions categories and attributes.
To address this gap, our EEmo-Bench utilizes an emotion ranking strategy to capture diverse emotional experiences and incorporate the VAD model to systematically evaluate MLLMs' ability to perceive and interpret emotions evoked by images.


\section{\textbf{EEmo-Bench} Dataset Construction} \label{sec: dataset construction}

\subsection{Basic principles}
The design of EEmo-Bench is guided by three core principles:
1) \textbf{Broad coverage of images}. Source images are collected from \textit{Flickr} \cite{FI-1, FI-2} with various content categories and are meticulously annotated by participants.
2) \textbf{Focus on evoked emotions}. Unlike previous benchmarks that emphasize expressed emotions conveyed by image \cite{MEMO-bench}, EEmo-Bench primarily targets the evoked emotions experienced by viewers \cite{Affection}. 
3) \textbf{Incorporation of pairwise comparisons}. To address the challenges of comparing emotional attributes evoked by image pairs, pairwise evaluation is integrated, enhancing the robustness of emotional understanding assessments.
The overall construction process is illustrated in Fig. \ref{fig:pipeline}.

\subsection{Emotional Attributes Definition}\label{sec: attributes}

For each image in EEmo-Bench, two types of emotional attributes are attached:
1) \textbf{Evoked emotions ranked by intensity}.
We employ Ekman's six basic emotions (joy, anger, disgust, sadness, surprise, and fear) along with `neutral' \cite{ekman1982emotion, Emotion6} to ensure clear emotion boundaries and minimize conflicts. Unlike existing benchmarks that focus predominantly on a single dominant emotion, we rank the top three evoked emotions from strongest to weakest. This approach accommodates individual preferences and cognitive variations by providing a more comprehensive and hierarchical emotional representation. 
2) \textbf{VAD scores}.
For a standardized and professional assessment, we utilize the VAD model, comprising three emotional attributes: 
a) Valence, which reflects the overall positive or negative emotion elicited by images, indicating the emotional tone of the viewer's response. 
b) Arousal, which represents the intensity of the evoked emotion, rating from intense to calm.
c) Dominance, which measures the degree of influence over
emotions and determines how emotions are experienced in terms of one’s sense of agency and authority in a particular emotional context, ranging from powerful to helpless.
For quantification, the Self-Assessment Manikin (SAM) 9-point scale \cite{IAPS} is adopted in EEmo-Bench, representing a standard and widely-used measure for assessing these attributes.
By integrating both emotion ranking and VAD scoring, EEmo-Bench offers a robust framework for evaluating the comprehensive emotional understanding capabilities of MLLMs.

\begin{figure*}
  \includegraphics[width=0.98\textwidth]{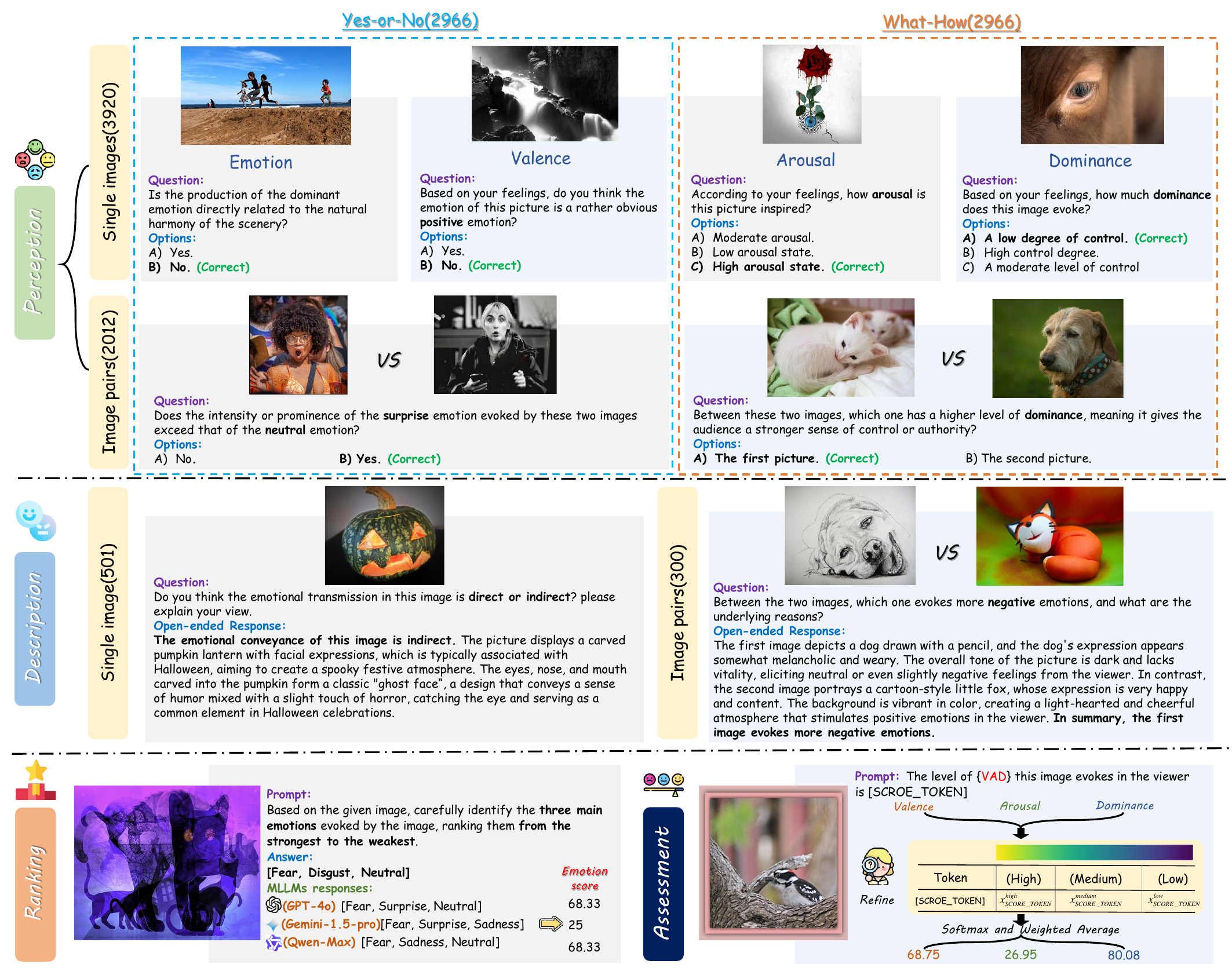}
  \vspace{-10pt}
  \caption{The visualization samples from \textbf{EEmo-Bench}, which involves four-dimensional tasks, including Perception, Ranking, Description, and Assessment. 
  }
  \Description{examples}
  \label{fig:examples}
  \vspace{-0.3cm}
\end{figure*}

\subsection{Source Images Collection} \label{sec:collection}

To ensure content diversity, the source images of EEmo-Bench are collected from  \textit{Flickr}, using the seven emotion categories outlined in Sec \ref{sec: attributes} as keywords for retrieving relevant images. 
Since the original labels often mix expressed emotions and evoked emotions, a manual annotation experiment is conducted to exclusively extract the evoked emotions, with details outlined in Sec. \ref{sec:annotation}.
In total, we gather $1,960$ images, with $280$ images for each emotion category. To further refine the dataset, content category classification is adopted, including animal, human, stationary object, daily life scene, natural landscape, and abstract/cartoon image.


\subsection{Subjective Annotation} \label{sec:annotation}

We conduct subjective experiments and recruit $15$ participants for manual annotation in strict compliance with the guidelines outlined in \cite{ITU-R} and \cite{IQA_protocol}. 
To accommodate participants unfamiliar with concepts like arousal and dominance, supplementary instructions (like tutorial videos and training sessions) are also provided. (\textit{See Supp. \ref{supp: annotation} for details}.)
Each participant is required to complete two tasks for every image:
1) Emotion selection and ranking. Choose up to three emotions from the seven candidate emotions and rank them by intensity from strongest to weakest.
2) VAD rating. Assess the valence, arousal, and dominance levels on the SAM $9$-point scale based on individual emotional response to the image.


After the annotation process, a total of $29,400$ labeled samples are collected. Each sample for the $i$-th image annotated by the $j$-th participant can be denoted as a quadruple 
 ($E, V, A, D$), where $E=[e_1, e_2, e_3]$ represents a list of emotions, sorted by intensity, with $E\in A_8^3 (e_{ANG}, e_{DIS}, e_{FEA}, e_{JOY}, e_{NEU}, e_{SAD}, e_{SUR},  None)$, and $V, A, D$ refers to the valence, arousal, and dominance, respectively. 
The emotions selected by participants are represented as $e_{EMO}$, where the emotion category can be abbreviated in subscript format.
To statistically sort the results of sentiment, we adopt a three-level criterion to calculate the rank of the three main evoked emotions, where the ranking score $S_{EMO}^i$ in the $i$-th image is denoted as:

\vspace{-10pt}
\begin{equation}
    \begin{split}
S&_{EMO}^i=W_{1}\sum_{k=1}^{3}\sum_{j=1}^{N}w_{k}\mathcal{R} (e_{ijk},e_{EMO}) \\
&+W_{2}\sum_{j=1}^{N}\mathcal{C} (e_{ijk},e_{EMO})
+W_{3}\sum_{k=1}^{3}\sum_{j=1}^{N}w_{i}^{\prime}\mathcal{R} (e_{ijk},e_{EMO}),
    \end{split}
    \label{equ:subject}
\end{equation}

\noindent where $N$ denotes the number of participants, $\mathcal{R} (e_{ijk},e_{EMO})$ is used to find if $e_{EMO}$ is the $k$-th ranked emotion of the $j$-th participant in the emotion ranking, and $\mathcal{C} (e_{ijk},e_{EMO})$ is adopted to determine whether $e_{EMO}$ is among the top three evoked emotions of $j$-th subject. 
By sorting the $S_{EMO}^i$ calculated by each emotion, the corresponding emotion rank could be acquired and served as ground truth. 
After several attempts, we set the weight coefficients of different positions as: $ (w_k=\{5,3,2\}$, $W_k=\{1000, 100, 10\}, w_{k}^{\prime}=\{1,0.1,0.01\}, k=1,2,3).$
The establishment of these weight coefficients forms a three-level comparison framework, where a small perturbation is introduced at each level to ensure that it breaks the tie when the preceding terms are equal, thus effectively avoids situations where the ranking scores are identical.
After ranking through these criteria, very few samples that still have conflicts are discarded, which has little impact on EEmo-Bench.

For the VAD scores, the method from \cite{Emotion6} is utilized, averaging the middle 9 scores from the 15 collected responses as the final result. 
Notably, we compute the standard deviation and maximum deviation from the mean for each image, excluding the other 6 scores. 
The maximum deviation is less than twice the standard deviation, demonstrating acceptable consistency for this inherently subjective task.
Fig. \ref{fig:pipeline} (b) displays the frequency distribution of the primary evoked emotions and the overall VAD score distribution.

Furthermore, the perception task Q\&A pairs in EEmo-Bench, as illustrated in Sec. \ref{sec: perception}, are designed based on these emotional attribute annotation results, without the need for additional manual labeling. 
For the description task, another four annotators are recruited to provide golden descriptions. (\textit{Detailed annotation guidelines can be found in Supp. \ref{supp: annotation}}.)

\section{EEmo-Bench Task Designs}

Based on the constructed dataset and emotional attributes, we propose \textbf{EEmo-Bench} to comprehensively evaluate the evoked emotion understanding abilities of multiple MLLMs from four tasks, including \textbf{Perception}, \textbf{Ranking}, \textbf{Description}, and \textbf{Assessment}. The evaluation samples are enumerated in Fig.\ref{fig:examples}.

\subsection{Perception} \label{sec: perception}

The perception task concentrates on evaluating the evoked emotion perception abilities of MLLMs, focusing on their accuracy in question answering related to emotion category and emotional attributes (see Fig. \ref{fig:pipeline} (c) (d)). 
The task can be categorized into the following three independent dimensions:

\noindent \textbf{Question types.} Two primary question types are engaged.
1) \textbf{Yes-or-No}: To address potential affirmative bias in MLLMs, the ratio of answers is balanced to approximately $1:1$ \cite{Q-bench-video}, ensuring the objectivity of the evaluation.
2) \textbf{What/How}: Designed to capture nuanced emotional perception, emphasizing more precise perception capabilities. While \textit{what} questions focus on identifying specific emotions and their causes, \textit{how} questions assess the perception of emotional properties, especially VAD attributes.

\begin{figure*}[htb]
\centering
\begin{subfigure}{0.28\textwidth}
    \centering
    \includegraphics[width=0.98\linewidth]{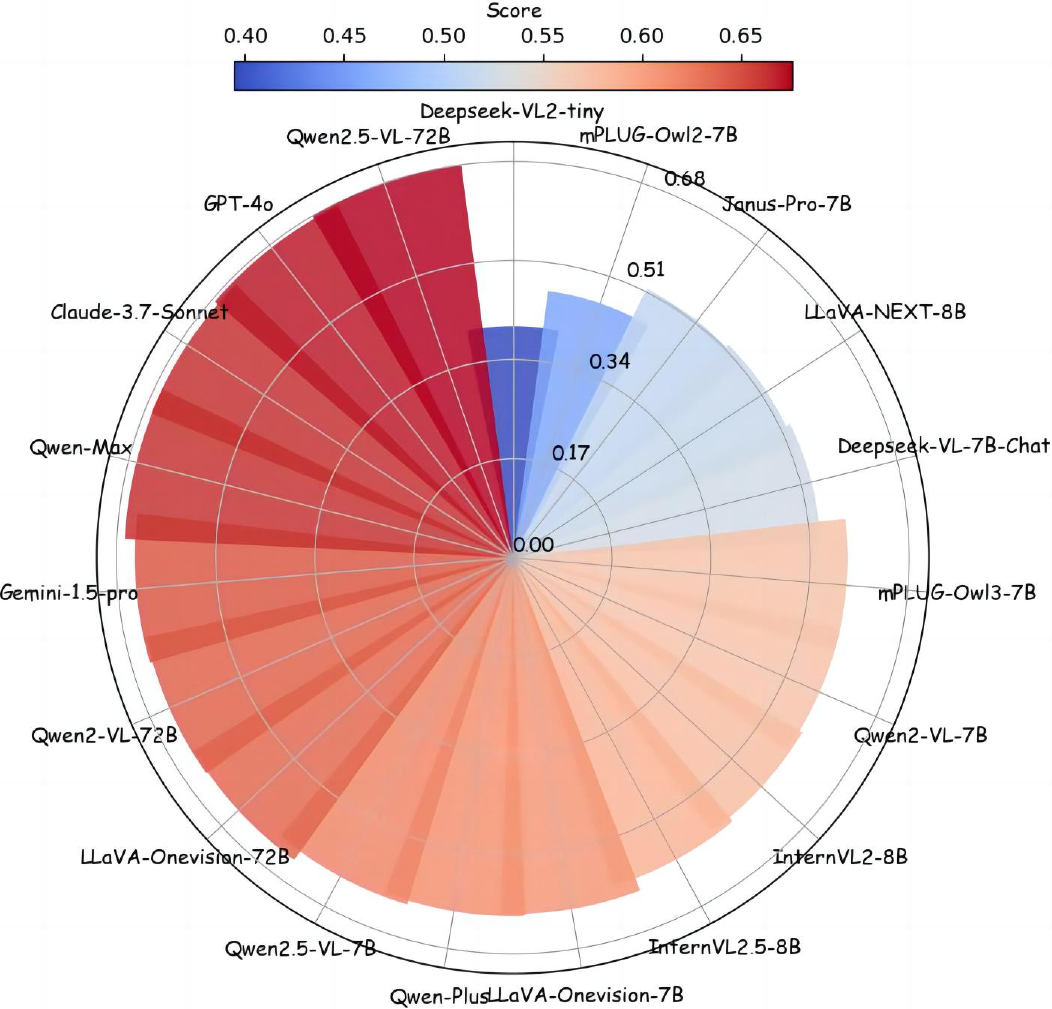}
    \caption{Overall performance on \textbf{EEmo-Bench}}
    \label{fig:overall} 
\end{subfigure}
\hfill
\begin{subfigure}{0.31\textwidth}
    \centering
    \includegraphics[width=\linewidth]{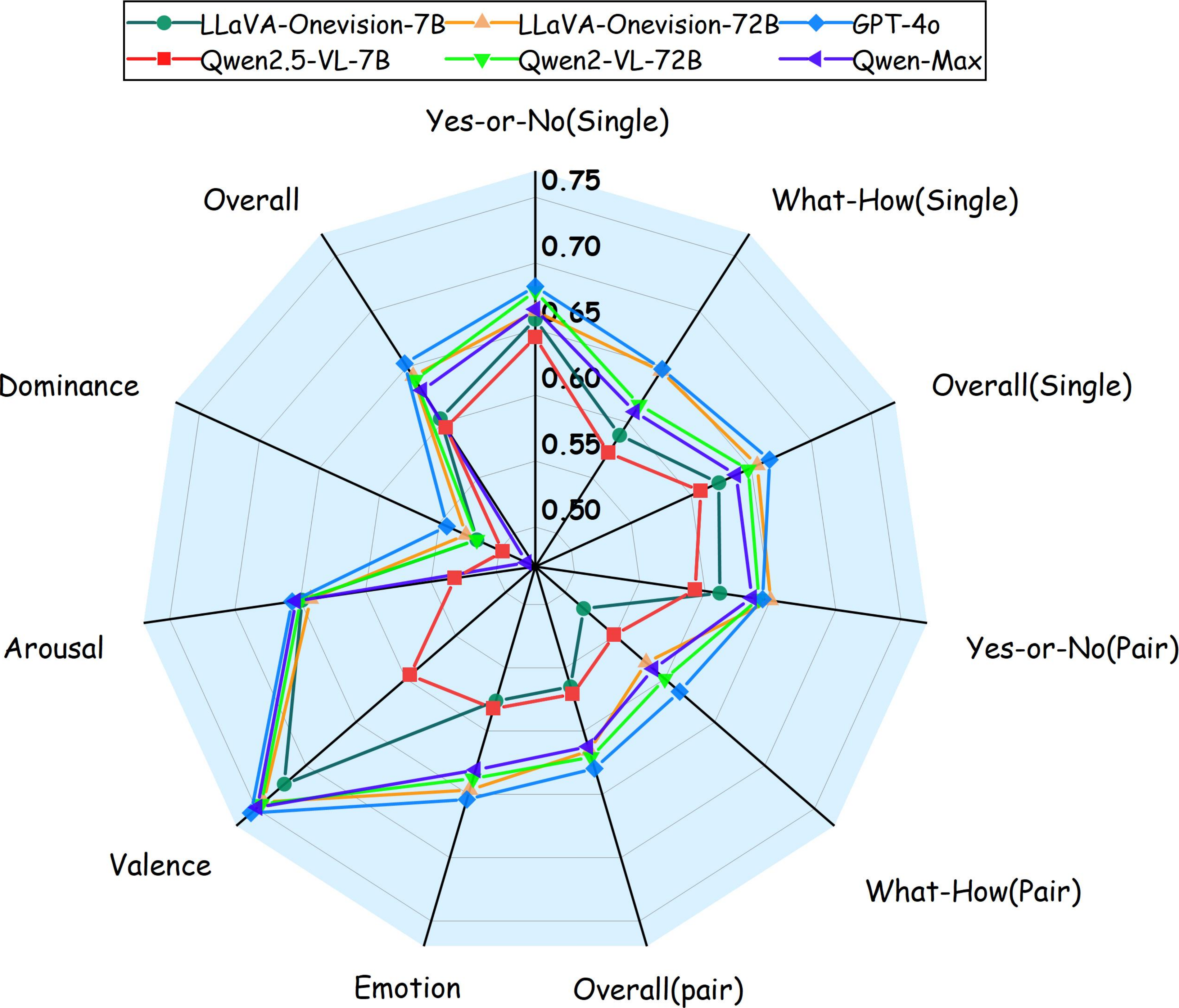}
    \caption{Performance on perception task}
    \label{fig:perception}
\end{subfigure}
\begin{subfigure}{0.38\textwidth}
    \centering
    \includegraphics[width=\linewidth]{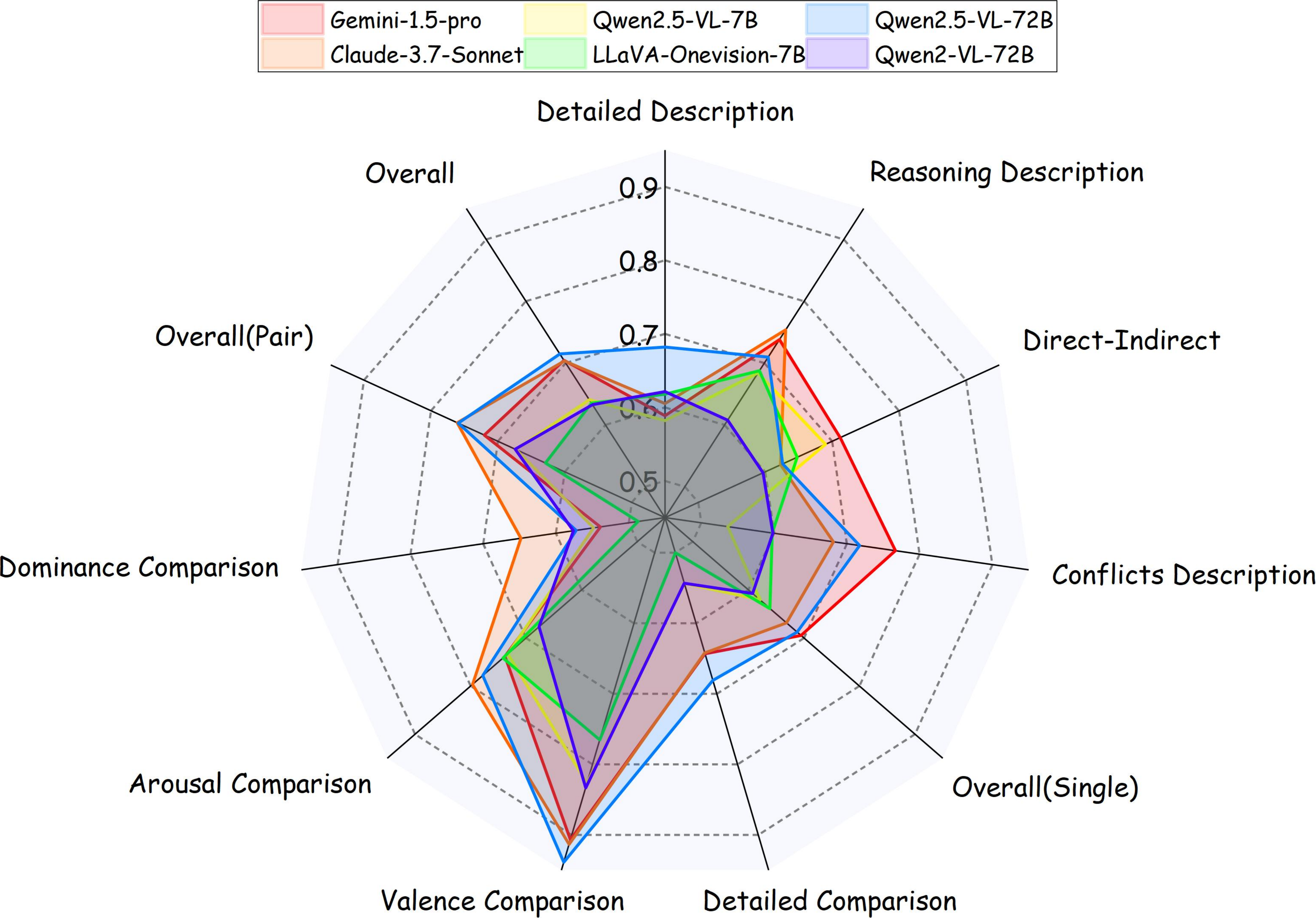}
    \caption{Performance on description task}
    \label{fig:description}
\end{subfigure}
\caption{A summary of the MLLM's performance on \textbf{EEmo-Bench}.
  (a) illustrates the overall performance of the $19$ involved MLLMs.
  (b) (c) displays comparisons of the \textbf{top-2} \textit{medium-scale open-source} MLLMs, \textit{large-scale open-source} MLLMs, and \textit{proprietary} MLLMs, throughout different dimensions in perception and description tasks, respectively.}
\label{fig: performance}
\vspace{-0.3cm}
\end{figure*}

\noindent \textbf{Single vs. Pairs.} \label{sec: single-pair}
Humans excel at perceiving emotions through comparative analysis of image pairs, particularly when assessing valence, arousal, and dominance. However, whether MLLMs can achieve similar accuracy in such comparative tasks remains uncertain. To address this, our benchmark is extended from single-image analysis to pairwise perception evaluation. (\textit{See Supp.\ref{supp: perception_pair_prompt} for detailed prompt}.)
 1) \textbf{Single images}.
Question concerns are divided into 
a) \textit{Emotion perception}: Focusing on evoked emotions extraction and VAD polarity understanding.
b) \textit{Reasoning}: Investigating the origin of emotions and the influence of specific factors.
These questions assess MLLMs' recognition and reasoning abilities, providing a comprehensive evaluation of their emotion perception from single images.
 2) \textbf{Image pairs}.
Question concerns are categorized into
a) \textit{Similarity extraction}: Identifying shared emotional properties, including emotion ranking, VAD polarity, and factors with common emotional impacts.
b) \textit{Differences comparison}:
Assessing discrepancies in emotion ranking and VAD scores across images.
To ensure meaningful comparisons, only image pairs with substantial VAD differences or opposite polarity and belonging to the same content category (e.g., both are animals) are selected.

\noindent \textbf{Perceptual dimensions.}
In Sec. \ref{sec: dataset construction}, we define four fundamental emotional attributes and conduct a subjective experiment to establish the ground truth for each image. EEmo-Bench designs perception questions across these dimensions, providing novel insights into MLLMs' emotional analysis capabilities. 
The dimensions are detailed as follows:
1) \textbf{Emotion}.
Unlike prior works \cite{Aesbench, UNIAA, MEMO-bench}  that focus solely on dominant emotions, EEmo-Bench examines the intensity relationships among multiple emotions evoked by one image and investigates the underlying reasons for each emotional response.
2) \textbf{Valence}.
As a core attribute in traditional emotion analysis, valence offers a broader assessment beyond specific emotion categories, capturing subtle shifts such as an image evoking peace and appreciation but slightly leaning toward negativity.
3) \textbf{Arousal}.
Emotions of the same category may exhibit slight differences due to variations in arousal levels.
For example, high-arousal joy is associated with excitement and satisfaction, while low-arousal joy is linked to calmness and relaxation. 
Understanding this distinction is essential for a nuanced emotional analysis.
4) \textbf{Dominance}.
Dominance is a more abstract emotional dimension, typically relying on an individual's emotional experience and cognitive judgment \cite{PAD}. This poses a significant challenge to the empathy capabilities of MLLMs, representing a critical yet underexplored aspect of comprehensive emotion perception.

\noindent \textbf{Evaluation process.} To address inconsistencies in MLLM outputs, 
we employ a $5$-round Deepseek-assisted evaluation strategy, which standardizes the outputs and enhances accurate assessment. For efficiency, consistent judgment results across repeated trials are reserved, while manual evaluations are required for the conflicting cases. (\textit{Details of the prompt are provided in Supp. \ref{supp: eval_perception}}.)

\subsection{Ranking} \label{sec: ranking}

This serves as a fundamental aspect of emotional attribute annotation and aims to assess the ability of MLLMs to detect relevant emotions and accurately rank them by intensity.
During each MLLM evaluation, the seven candidate emotions are provided as options, and we instruct the MLLMs to choose and rank no more than three predominant evoked emotions according to their intensity.

\noindent \textbf{Evaluation process:} Based on the result provided by MLLMs and the ground truth, we employ a strategy to calculate the \textbf{emotion ranking score} to assess the performance. 
The rank of the emotions evoked in the $i$-th image is denoted as 
$E_{GT}^i=[e_1^i, e_2^i, e_3^i]$, and the predicted rank by each MLLM is represented as $E_{MLLM}^i=[e_1^{i\prime}, e_2^{i\prime}, e_3^{i\prime}]$, with $E_{GT}^i,E_{MLLM}\in A_8^3 (e_{ANG}, e_{DIS}, e_{FEA}, e_{JOY}, e_{NEU}, e_{SAD}, e_{SUR}, \\None)$. The ranking score $S_{RANK}^i$
of each image is shown as follows:

\begin{equation}
    \begin{split}
        S_{RANK}^{i}&= (\sum_{k=1}^{3}\sum_{j=1}^{3}w_{k}\mathcal{R} (e_{k}^{i},e_{j}^{i\prime}))_{Scaled} \\
    &+\mathcal{W} (E_{MLLM}^{i},E_{GT}^{i})\cdot (\mathcal{K} (\mathcal{SP} (E_{MLLM}^{i},E_{GT}^{i}))_{Scaled},
    \end{split}
\end{equation}

\noindent where  $\mathcal{R} (e_{k}^{i},e_{j}^{i})$ and $\{w_{k}| k=1,2,3\}$
share the definition in Equ.\ref{equ:subject},
$ (\cdot)_{Scaled}$ stands for the scaled function to map the value to the range of $50$, 
$\mathcal{SP} (\cdot)$ indicates the function to sample the common elements between two emotion lists while preserving their original order,
$\mathcal{K} (\cdot)$ represents Kendall's Tau function, which is used to measure the ranking performance for small sample sizes, 
and $\mathcal{W} (\cdot)$ denotes a weight function based on the length of the predicted list ($1, \frac{1}{3},0$ for the length of $1,2,3$).
After that, $S_{RANK}^i$ of all images is averaged, representing the final ranking result.

\subsection{Description}

We also assess the descriptive abilities of MLLMs. 
Questions are designed to evaluate comprehensive understanding capabilities based on two categories:
1) \textbf{Single image analysis}. This involves assessing capabilities in detailed analysis, reasoning, direct/indirect emotion evocation, and explaining conflicting emotions.
2) \textbf{Image pair comparison}. This focuses on evaluating the ability to compare emotional properties, including recognizing similarities and differences in emotion ranking and differentiating VAD polarity between two images.
To evaluate MLLM performance, we compare their outputs with the golden descriptions using a $5$-round Deepseek-assisted voting as well, measuring  \textbf{Completeness, Accuracy, and Relevance}. (\textit{Details can be found in Supp. \ref{supp: eval_description}}.)

\begin{table*}[t]
\footnotesize
\setlength{\tabcolsep}{4pt} 
\renewcommand{\arraystretch}{0.85}   
\caption{Result on the image emotions perception across single images and image pairs and the emotion ranking abilities of MLLMs. The best performance is \textbf{bolded} and the second and third performances are {\ul underlined.}}
\vspace{-6pt} 
\begin{tabular}{lccc|ccc|cccc|c|c}
\hline
\multicolumn{1}{c|}{\textbf{Tasks}} & \multicolumn{11}{c|}{\textbf{Perception}}  & \multicolumn{1}{c}{\textbf{Ranking}} \\ \hline
\multicolumn{1}{c|}{\textbf{Sub-categories}} & \multicolumn{3}{c|}{\textbf{Single Image}}  & \multicolumn{3}{c|}{\textbf{Image Pair}}  & \multicolumn{4}{c|}{\textbf{Perceptual Dimensions}}  &    &    \\ \cline{1-11}
\multicolumn{1}{l|}{{\scriptsize \textbf{MLLMs}}} & {\scriptsize \textit{Yes-or-No↑}} & {\scriptsize \textit{What-How↑}}& {\scriptsize \textit{Overall↑}} & {\scriptsize \textit{Yes-or-No↑}} & {\scriptsize \textit{What-How↑}}& {\scriptsize \textit{Overall↑}} & {\scriptsize \textit{Emotion↑}}   & {\scriptsize \textit{Valence↑}}   & {\scriptsize \textit{Arousal↑}}   & {\scriptsize \textit{Dominance↑}} & \multirow{-2}{*}{{\scriptsize \textit{Overall↑}}} & \multirow{-2}{*}{\begin{tabular}[c]{@{}c@{}}{\scriptsize \textit{Emotion}} \\ {\scriptsize \textit{Score↑}} \end{tabular}}       \\ \hline
\multicolumn{1}{l|}{{\scriptsize \textbf{Random guess}}}  & 50.00\%& 33.33\%& 41.67\%& 50.00\%& 34.26\%& 42.13\%& 39.38\% & 44.62\% & 44.72\% & 44.44\% &  41.83\% & 25.48\%\\ \hline
\textcolor{gray}{\scriptsize \textit{Medium-scale open-source MLLMs}}\\  \hdashline 
\multicolumn{1}{l|}{{\scriptsize Deepseek-VL2-tiny \cite{deepseek-vl2}}}   & 54.62\%          & 46.99\%          & 50.81\%          & 50.40\%          & 38.41\%          & 44.41\%          & 45.47\%          & 53.97\%          & 53.99\%          & 48.13\%          & 48.63\%                  & 21.82\%                        \\
\multicolumn{1}{l|}{{\scriptsize Deepseek-VL-7B-Chat \cite{deepseek-vl}}} & 60.18\%          & 50.14\%          & 55.16\%          & 52.88\%          & 38.71\%          & 45.80\%          & 51.30\%          & 59.41\%          & 56.36\%          & 43.93\%          & 51.98\%                  & 57.12\%                        \\
\multicolumn{1}{l|}{{\scriptsize InternVL2-8B \cite{internvl2}}}   & 58.04\%          & 53.62\%          & 55.83\%          & 57.06\%          & 49.45\%          & 53.26\%          & 55.16\%          & 62.55\%          & 54.09\%          & 47.11\%          & 54.96\%                  & 57.07\%                        \\
\multicolumn{1}{l|}{{\scriptsize InternVL2.5-8B \cite{Internvl2.5}}}  & 59.85\%          & 54.59\%          & 57.22\%          & 60.74\%          & 54.93\%          & 57.84\%          & 55.34\%          & 65.27\%          & 60.78\%          & 51.99\%          & 57.43\%                  & 57.87\%                        \\
\multicolumn{1}{l|}{{\scriptsize Janus-Pro-7B \cite{janus}}}    & 57.53\%          & 48.83\%          & 53.18\%          & 53.18\%          & 43.48\%          & 48.33\%          & 50.20\%          & 56.28\%          & 54.53\%          & \textbf{57.67\%} & 51.53\%                  & 54.83\%                        \\
\multicolumn{1}{l|}{{\scriptsize  LLaVA-Onevision-7B \cite{LLava-onevison}}}   & 65.75\%          & 58.83\%          & 62.29\%          & 61.13\%          & 51.84\%          & 56.49\%          & 57.62\%          & 72.18\%          & 64.87\%          & 51.87\%          & 60.32\%                  & 58.39\%                        \\
\multicolumn{1}{l|}{{\scriptsize  LLaVA-NEXT-8B \cite{LLava-next}}}  & 57.43\%          & 55.87\%          & 56.65\%          & 55.77\%          & 43.18\%          & 49.48\%          & 53.13\%          & 63.39\%          & 53.34\%          & 47.90\%          & 54.22\%                  & 54.55\%                        \\
\multicolumn{1}{l|}{{\scriptsize mPLUG-Owl2-7B \cite{mplug2}}}    & 61.00\%          & 50.31\%          & 55.66\%          & 60.83\%          & 38.21\%          & 49.52\%          & 50.83\%          & 65.38\%          & 58.73\%          & 45.40\%          & 53.57\%                  & 48.26\%                        \\
\multicolumn{1}{l|}{{\scriptsize mPLUG-Owl3-7B \cite{mplug3}}}     & 59.11\%          & 52.96\%          & 56.04\%          & 59.84\%          & 52.84\%          & 56.34\%          & 54.06\%          & 67.89\%          & 58.41\%          & 48.01\%          & 56.14\%                  & 58.53\%                        \\
\multicolumn{1}{l|}{{\scriptsize Qwen2-VL-7B \cite{qwen2}}}     & 59.98\%          & 59.39\%          & 59.69\%          & 54.97\%          & 54.43\%          & 54.70\%          & 58.62\%          & 66.42\%          & 56.36\%          & 48.24\%          & 57.99\%                  & 61.36\%                        \\
\multicolumn{1}{l|}{{\scriptsize Qwen2.5-VL-7B \cite{qwen2.5}}}     & 64.37\%          & 57.24\%          & 60.81\%          & 59.24\%          & 54.93\%          & 57.09\%          & 58.22\%          & 59.56\%          & 53.15\%          & 49.72\%          & 59.54\%                  & 61.88\%                  \\ \hline
\textcolor{gray}{\scriptsize \textit{Large-scale open-source MLLMs}}\\  \hdashline 
\multicolumn{1}{l|}{{\scriptsize  LLaVA-Onevision-72B \cite{LLava-onevison}}}     & 66.36\%          & {\ul 64.64\%}    & {\ul 65.50\%}    & \textbf{65.01\%} & 58.11\%          & 61.56\%          & {\ul 64.65\%}    & 74.27\%          & 64.22\%          & {\ul 52.78\%}    & {\ul 64.16\%}            & 66.92\%                             \\
\multicolumn{1}{l|}{{\scriptsize Qwen2-VL-72B \cite{qwen2}}}     & {\ul 67.89\%}    & {\ul 61.58\%}    & {\ul 64.74\%}    & {\ul 64.12\%}    & {\ul 60.00\%}    & {\ul 62.06\%}    & 63.75\%          & 74.48\%          & 64.98\%          & 51.87\%          & {\ul 63.83\%}            & 64.69\%                             \\
\multicolumn{1}{l|}{{\scriptsize Qwen2.5-VL-72B \cite{qwen2.5}}}    & {\ul 67.08\%}    & 61.28\%          & 64.18\%          & 63.92\%          & 59.80\%          & {\ul 61.86\%}    & {\ul 63.81\%}    & {\ul 75.00\%}    & {\ul 65.30\%}    & 48.92\%          & 63.39\%                  & \textbf{67.84\%}                            \\ \hline
\textcolor{gray}{\scriptsize \textit{Proprietary MLLMs}}\\  \hdashline 
\multicolumn{1}{l|}{{\scriptsize Gemini-1.5-pro \cite{gemini}}}   & 67.06\%          & 59.65\%          & 63.36\%          & 61.00\%          & 58.27\%          & 59.64\%          & 60.23\%          & \textbf{76.02\%} & 63.00\%          & 52.44\%          & 62.09\%                  & 60.65\%                        \\
\multicolumn{1}{l|}{{\scriptsize GPT-4o \cite{GPT4}}}   & \textbf{68.25\%} & \textbf{64.80\%} & \textbf{66.53\%} & {\ul 64.41\%}    & \textbf{61.49\%} & \textbf{62.95\%} & \textbf{65.41\%} & {\ul 75.52\%}    & \textbf{65.62\%} & {\ul 54.37\%}    & \textbf{65.31\%}         & 65.67\%                  \\
\multicolumn{1}{l|}{{\scriptsize Qwen-VL-Max \cite{qwen-vl}}}   & 66.51\%          & 60.96\%          & 63.74\%          & 63.59\%          & 58.84\%          & 61.22\%          & 63.06\%          & 74.87\%          & {\ul 65.33\%}    & 47.71\%          & 62.88\%                  & {\ul 67.27\%}               \\
\multicolumn{1}{l|}{{\scriptsize Qwen-VL-Plus \cite{qwen-vl}}}    & 63.56\%          & 55.69\%          & 59.63\%          & 62.39\%          & 55.72\%          & 59.06\%          & 57.64\%          & 68.17\%          & 62.43\%          & 52.06\%          & 59.43\%                  & 61.19\%                        \\
\multicolumn{1}{l|}{{\scriptsize Claude-3.7-Sonnet \cite{anthropic2025claude}}}   & 64.08\%          & 59.04\%          & 61.56\%          & 61.93\%          & \textbf{61.49\%} & 61.71\%          & 60.69\%          & 73.82\%          & 61.42\%          & 52.05\%          & 61.61\%                  & {\ul 67.05\%}                   \\    \hline
\end{tabular}
\label{tab: perception}
\vspace{-0.3cm}
\end{table*}

\subsection{Assessment} \label{sec: assessment}

In the final task, we benchmark the ability of MLLMs to quantify the \textbf{VAD} scores.
Inspired by the zero-shot quantitative scoring methodology in \cite{Q-bench, Q-BENCH*, Q-Eval-100K}, emotional level keywords are defined for each VAD attribute: {\textit{Positive, Neutral, Negative}} for \textbf{Valence}; {\textit{High, Moderate, Low}} for \textbf{Arousal}; and {\textit{Powerful, Moderate, Helpless}} for \textbf{Dominance}, which can be mapped to a unified scale of {\textit{High, Medium, Low}}.
We apply softmax pooling on the logits of extracted keywords to derive probabilistic scores for each rating level:

\begin{equation}
    \vspace{-2pt}
    p_{l}=\frac{e^{x_{SCORE\_TOKEN}^l}}{\sum_{l}^{l\in\mathcal{L}}e^{x_{SCORE\_TOKEN}^l}},
\end{equation}

\noindent where $\mathcal{L}$ indicates the keywords set of levels (\textit{High, Medium, Low}), and $x_{SCORE\_TOKEN}^l$, $p_{l}$ represent the logits and the probabilities of keywords in different levels, respectively. 
Finally, we determine the final predicted rating $r_{VAD}$ using weighted average of $p_{l}$:

\begin{equation}
    r_{VAD} = \sum_{l}^{l\in\mathcal{L}}w_l^r\cdot p_l,
\end{equation}

\noindent where $w_l^r$ is the numerical weight for each rating level (with $w_l^r=\{1, 0.5, 0\}$ corresponding to the levels of \textit{High}, \textit{Medium}, and \textit{Low}, respectively.). For evaluation, EEmo-Bench calculates the correlation values, including Spearman's Rank Correlation Coefficient (SRCC) and Pearson Linear Correlation Coefficient (PLCC), between $r_{VAD}$ and the related golden scores.

\section{Experiment}

\subsection{Experiment Setup}

\sloppy
In this section, we evaluate the performance of $5$ \textit{proprietary} MLLMs and $14$ \textit{open-source} MLLMs, while the latter are divided by parameter scale into two categories: medium-scale (lower than 8b parameters) and large-scale (72b parameters).
To be specific, MLLMs are classified into
1) \textit{Proprietary MLLMs}: Gemini-1.5-pro \cite{gemini}, GPT-4o \cite{GPT4}, Qwen-VL-Plus, Qwen-VL-Max \cite{qwen-vl} and Claude-3.7-Sonnet \cite{anthropic2025claude}. 
2) \textit{Medium-scale Open-source MLLMs}: Deepseek-VL-7B-Chat \cite{deepseek-vl}, Deepseek-VL2-tiny \cite{deepseek-vl2}, InternVL2-8B \cite{internvl2}, InternVL2.5-8B \cite{Internvl2.5}, Janus-Pro-7B \cite{janus}, LLaVA-Onevision (Qwen2-7B) \cite{LLava-onevison}, LLaVA-NeXT (Llama3-8B) \cite{LLava-next}, mPLUG-Owl2 (LLaMA2-7B) \cite{mplug2}, mPLUG-Owl3 (Qwen2-7B) \cite{mplug3},
Qwen2-VL-7B-Instruct \cite{qwen2}, and Qwen2.5-VL-7B-Instruct \cite{qwen2.5}.
3) \textit{Large-scale Open-source MLLMs}: LLaVA-Onevision (Qwen2-72B) \cite{LLava-onevison}, Qwen2-VL-72B \cite{qwen2}, and Qwen2.5-VL-72B \cite{qwen2.5}.

\begin{table*}[t]
\footnotesize
\setlength{\tabcolsep}{5.5pt} 
\renewcommand{\arraystretch}{0.85}   
\caption{Result on the evoked emotions description abilities across single images and image pairs of MLLMs. The best performance is \textbf{bolded}, and the second and third performances are {\underline{underlined}.}}
\vspace{-6pt}
\begin{tabular}{lccccc|c@{\hspace{1pt}}c@{\hspace{1pt}}c@{\hspace{1pt}}c@{\hspace{1pt}}c|c}
\hline
\multicolumn{1}{c|}{Sub-categories} & \multicolumn{5}{c|}{Single}  & \multicolumn{5}{c|}{Pair} & \\ \cline{1-11}
\multicolumn{1}{l|}{{\scriptsize \textbf{MLLMs}}} & \begin{tabular}[c]{@{}c@{}}{\scriptsize \textit{Detailed}}\\ {\scriptsize \textit{Description↑}}\end{tabular} & \begin{tabular}[c]{@{}c@{}}{\scriptsize \textit{Reasoning}}\\ {\scriptsize \textit{Description↑}}\end{tabular} & \begin{tabular}[c]{@{}c@{}}{\scriptsize \textit{Direct}}\\ {\scriptsize \textit{-Indirect↑}}\end{tabular} & \begin{tabular}[c]{@{}c@{}}{\scriptsize \textit{Conflicts}}\\ {\scriptsize \textit{Description↑}}\end{tabular} & {\scriptsize \textit{Overall↑}}         & \begin{tabular}[c]{@{}c@{}}{\scriptsize \textit{Detailed}}\\ {\scriptsize \textit{Comparison↑}}\end{tabular} & \begin{tabular}[c]{@{}c@{}}{\scriptsize \textit{Valence}}\\ {\scriptsize \textit{Comparison↑}}\end{tabular} & \begin{tabular}[c]{@{}c@{}}{\scriptsize \textit{Arousal}}\\ {\scriptsize \textit{Comparison↑}}\end{tabular} & \begin{tabular}[c]{@{}c@{}}{\scriptsize \textit{Dominance}}\\ {\scriptsize \textit{Comparison↑}}\end{tabular} & {\scriptsize \textit{Overall↑}}  & \multirow{-2}{*}{{\scriptsize \textit{Overall↑}}} \\ \hline
\textcolor{gray}{\scriptsize \textit{Medium-scale open-source MLLMs}}\\  \hdashline 
\multicolumn{1}{l|}{{\scriptsize Deepseek-VL2-tiny} }        & 50.07\%  & 54.76\% & 53.91\% & 44.30\% & 50.82\% & 32.91\% & 56.93\% & 44.67\% & 36.48\%  & 43.13\% & 47.94\%                   \\
\multicolumn{1}{l|}{{\scriptsize Deepseek-VL-7B-Chat}}       & 45.47\%  & 53.66\% & 55.11\% & 44.82\% & 49.46\% & 39.75\% & 58.13\% & 48.67\% & 33.33\%  & 46.73\% & 48.44\%                   \\
\multicolumn{1}{l|}{{\scriptsize InternVL2-8B}}   & 56.07\%  & 65.79\% & 64.67\% & 53.95\% & 59.98\% & 51.65\% & 74.00\% & 66.00\% & 54.81\%  & 62.73\% & 61.01\%                   \\
\multicolumn{1}{l|}{{\scriptsize InternVL2.5-8B}} & 55.53\%  & 67.52\% & 61.74\% & 53.60\% & 59.70\% & 49.62\% & 75.47\% & 66.17\% & 54.67\%  & 61.30\% & 60.30\%                   \\
\multicolumn{1}{l|}{{\scriptsize Janus-Pro-7B}}   & 54.67\%  & 59.66\% & 44.35\% & 41.23\% & 50.26\% & 40.13\% & 56.00\% & 36.33\% & 32.96\%  & 42.27\% & 47.27\%                   \\
\multicolumn{1}{l|}{{\scriptsize LLaVA-Onevision-7B}}       & 61.80\%  & 68.76\% & 64.78\% & 59.74\% & 63.89\% & 50.00\% & 76.53\% & 74.17\% & 48.70\%  & 62.90\% & 63.52\%                   \\
\multicolumn{1}{l|}{{\scriptsize LLaVA-NEXT-8B}}  & 50.00\%  & 50.76\% & 56.74\% & 53.42\% & 52.24\% & 29.49\% & 43.47\% & 32.83\% & 36.30\%  & 36.57\% & 46.37\%                   \\
\multicolumn{1}{l|}{{\scriptsize mPLUG-Owl2-7B}}  & 44.27\%  & 47.66\% & 41.20\% & 38.42\% & 43.35\% & 23.42\% & 20.93\% & 25.50\% & 17.59\%  & 22.87\% & 35.68\%                   \\
\multicolumn{1}{l|}{{\scriptsize mPLUG-Owl3-7B}}  & 54.87\%  & 63.24\% & 52.72\% & 45.44\% & 54.75\% & 46.46\% & 74.53\% & 65.17\% & 51.11\%  & 59.47\% & 56.52\%                   \\
\multicolumn{1}{l|}{{\scriptsize Qwen2-VL-7B}}    & 55.93\%  & 58.76\% & 50.00\% & 29.30\% & 49.60\% & 43.29\% & 72.40\% & 57.50\% & 47.96\%  & 55.87\% & 51.95\%                   \\
\multicolumn{1}{l|}{{\scriptsize Qwen2.5-VL-7B}}  & 58.20\%  & 68.41\% & {\ul 69.02\%}     & 53.60\% & 62.10\% & 54.30\% & 82.67\% & 73.33\% & 54.81\%  & 67.33\% & 64.06\%                   \\ \hline
\textcolor{gray}{\scriptsize \textit{Large-scale open-source MLLMs}}\\  \hdashline 
\multicolumn{1}{l|}{{\scriptsize LLaVA-Onevision-72B}}       & 51.13\%  & 61.79\% & 49.46\% & 63.77\% & 56.79\% & 47.34\% & 82.93\% & 76.33\% & 47.78\%  & 64.73\% & 59.76\%                   \\
\multicolumn{1}{l|}{{\scriptsize Qwen2-VL-72B}}   & {\ul 62.13\%} & 60.76\% & 59.67\% & 59.82\% & 60.76\% & 54.30\% & 83.33\% & 67.67\% & {\ul 57.59\%} & 67.43\% & 63.26\%                   \\
\multicolumn{1}{l|}{{\scriptsize Qwen2.5-VL-72B}} & \textbf{68.2\%}       & 70.97\% & 62.61\% & {\ul 71.75\%} & {\ul 68.78\%}    & \textbf{68.10\%}     & \textbf{93.87\%}    & 77.83\%       & 57.22\% & {\ul 75.93\%} & \textbf{71.46\%}          \\ \hline
\textcolor{gray}{\scriptsize \textit{Proprietary MLLMs}}\\  \hdashline 
\multicolumn{1}{l|}{{\scriptsize Gemini-1.5-pro}} & 58.87\%  & {\ul 73.79\%}       & \textbf{71.20\%}  & \textbf{76.67\%}       & \textbf{69.50\%} & {\ul 64.30\%}        & {\ul 90.67\%}       & 73.83\% & 53.89\%  & 72.03\% & {\ul 70.45\%}             \\
\multicolumn{1}{l|}{{\scriptsize GPT-4o}}        & {\ul 63.93\%} & 71.03\% & {\ul 69.24\%}     & 66.93\% & {\ul 67.64\%}    & 56.96\% & {\ul 90.67\%}       & {\ul 79.33\%}       & {\ul 64.44\%}      & {\ul 73.93\%}    & 70.00\%             \\
\multicolumn{1}{l|}{{\scriptsize Qwen-VL-Max}}       & 60.14\%  & {\ul 71.86\%} & 61.65\% & {\ul 68.42\%} & 65.73\% & 62.15\%        & 88.51\% & \textbf{80.85\%}    & 55.85\%  &  73.06\%    & 68.48\%                   \\
\multicolumn{1}{l|}{{\scriptsize Qwen-VL-Plus}}      & 60.27\%  & 63.72\% & 66.70\% & 54.65\% & 61.17\% & 53.80\% & 80.14\% & 71.69\% & 49.43\%  & 65.35\% & 62.74\%                   \\ 
\multicolumn{1}{l|}{{\scriptsize Claude-3.7-Sonnet}}      & 60.53\%  & \textbf{75.38\% }& 62.28\% & 68.16\% & 66.89\% & {\ul 64.18\%} & {\ul 91.33\%} & {\ul 79.67\%} & \textbf{64.81\%}  & \textbf{76.10\%} & {\ul 70.34\%}                   \\ \hline
\end{tabular}
\label{tab: description}
\vspace{-0.3cm}
\end{table*}

\subsection{Observations}
\label{sec: observation}

The overall performance on \textbf{EEmo-bench} and subcategories comparison on the perception and description task is shown in Fig. \ref{fig: performance}. The observations on each task are illustrated as follows:

\noindent{\textbf{Result of Perception Task.}}
The results presented in Tab. \ref{tab: perception} indicate that all MLLMs significantly surpass the random guessing results in perception tasks, confirming their basic competence in evoked emotions perception. Among these, GPT-4o achieves the highest overall performance at $65.31\%$, closely followed by large-scale open-source models such as LLaVA-Onevision-72B and Qwen2-VL-72B, scoring $64.16\%$ and $63.83\%$, respectively. 
Notably, large-scale open-source MLLMs generally outperform the proprietary MLLMs (except GPT-4o), underscoring their capacity in image emotion perception tasks. Despite their performance, even the best models fall short of exceptional emotional perception standards, revealing considerable room for improvement in evoked emotions understanding. Detailed observations are provided from the following perspectives.


1) Single vs. Pairs.
Results in Tab. \ref{tab: perception} demonstrate that MLLMs' average performance for single-image tasks exceeds paired-image tasks by approximately $\textbf{3.7\%}$, despite a slightly lower random-guess baseline. 
This suggests that MLLMs have stronger capabilities for emotional perception and reasoning tasks within individual images but face challenges when encountering double-stimulus analysis.
(\textit{The analysis of the four dimensions, segmented by question concerns, is provided in Supp. \ref{supp: extended results}.})

2) Perceptual dimensions.
Further analysis from Tab. \ref{tab: perception} reveals that MLLMs effectively recognize salient emotional features, such as valence that derived from colors and representative objects. However, their understanding remains insufficient for more nuanced emotional attributes, like dominance and arousal, that require a deeper integration of human common sense and subjective preferences. This underscores the need to enhance MLLMs' systematic capabilities regarding emotions evoked by images. (\textit{We further illustrate our findings on $6$ content categories in Supp. \ref{supp: extended results}}.)

\noindent{\textbf{Result of Ranking Task.}}
As presented in Tab. \ref{tab: perception}, most MLLMs achieve basic performance in ranking the evoked emotions, significantly surpassing random guesses. 
Among that, Qwen2.5-VL-72B exhibits the highest score at $67.84\%$, closely followed by Qwen-VL-Max and Claude-3.7-Sonnet at $67.27\%$ and $67.05\%$, respectively. 
However, even these top-performing MLLMs still exhibit a considerable gap compared to human perception.
Although capable of identifying primary evoked emotions in many cases, their limited sensitivity to emotional intensity often leads to inconsistent or reversed emotion ranking, ultimately diminishing the overall score. These results underscore the necessity of further enhancing MLLMs’ capabilities in comprehensively perceiving emotional intensity, which is crucial for assessing the variability in emotional experiences across different individuals as well as the diverse emotional responses of a single individual based on images.

\noindent{\textbf{Result of Description Task.}} 
Based on the results presented in Tab. \ref{tab: description}, we summarize our key findings as follows:

1) Overall performance.
Qwen2.5-VL-72B achieves the highest overall proficiency, indicating that large-scale open-source models have reached a promising level in detailed emotional analysis. 
However, proprietary models, notably Gemini-1.5-pro and Claude-3.7-Sonnet, significantly outperform other open-source models on open-ended emotional questions. 
This advantage is attributed to their more comprehensive and extensive reasoning capabilities, aligning closely with human emotional understanding processes.

2) Unbalanced performance across perspectives.
MLLMs demonstrate superior performance in comparative tasks involving image pairs but exhibit significant variability across subtasks that focus on different emotional attributes. 
Specifically, MLLMs demonstrate strong performance on valence, arousal, and reasoning subtasks, with Qwen2.5-VL-72B achieving an excellent performance of $93.87\%$ on valence comparisons, fully meeting daily evaluation requirements for this task. 
This highlights their strengths in recognizing typical emotional cues derived from symbolic objects and color schemes. 
Nevertheless, notable shortcomings remain in their nuanced perception and detailed comparison of emotional diversity, particularly regarding dominance. 
Thus, most MLLMs possess a limited and superficial understanding of image-evoked emotions, underscoring the gap relative to human emotional comprehension.

\begin{table}[t]
\centering
\renewcommand\arraystretch{1}
\footnotesize
\setlength{\tabcolsep}{5.5pt} 
\renewcommand{\arraystretch}{0.85}   
\caption{Result for the Valence-Arousal-Dominance (VAD) assessment ability of the evoked emotions of several medium-scale open-source MLLMs. Metrics are \textit{SRCC↑}/\textit{PLCC↑}, with the best performance \textbf{bolded}.}
\vspace{-5pt}
\resizebox{1\linewidth}{!}
{\begin{tabular}{l|ccc|c}
\hline
\multicolumn{1}{c|}{\textbf{Dimensions/Model}} & \textbf{Valence}  & \textbf{Arousal}  & \textbf{Dominance}  & \textbf{Overall}  \\ \hline
{\scriptsize Deepseek-VL2-tiny}  & 0.81~/~0.78 & 0.43~/~0.40 & -0.18~/~-0.18 & 0.35~/~0.33 \\
{\scriptsize Deepseek-VL-7B-Chat} & 0.85~/~0.83 & 0.54~/~0.56 & -0.50~/~-0.53 & 0.30~/~0.29 \\
{\scriptsize Janus-Pro-7B}       & 0.76~/~0.76 & 0.54~/~0.38 & -0.30~/~-0.23 & 0.33~/~0.30 \\
{\scriptsize LLaVA-Onevision-7B} & 0.67~/~0.47 & 0.24~/~0.28 &  -0.19~/~-0.17  & 0.24~/~0.19 \\
{\scriptsize LLaVA-NEXT-8B}      & 0.58~/~0.57 & 0.38~/~0.37 & -0.07~/~-0.05   & 0.30~/~0.30 \\
{\scriptsize mPLUG-Owl3-7B}      & 0.85~/~0.86 & 0.56~/~0.53 & -0.22~/~-0.07 & 0.40~/~0.44 \\
{\scriptsize Qwen2-VL-7B}        & \textbf{0.87~/~0.88} & \textbf{0.59~/~0.57} & -0.04~/~-0.04 & 0.47~/~0.47 \\
{\scriptsize Qwen2.5-VL-7B}      & 0.80~/~0.74 & 0.46~/~0.46 & \textbf{0.32~/~0.31}   & \textbf{0.53~/~0.50} \\ \hline
\end{tabular}}
\label{tab:assessment}
\vspace{-0.15cm}
\end{table}

\noindent{\textbf{Result of Assessment Task.}}
From Tab. \ref{tab:assessment}, we observe robust performance for valence, averaging around $\textbf{0.77}$, and moderate performance for arousal, averaging approximately $\textbf{0.46}$. This finding further confirms their capability in coarse-grained emotional comprehension, particularly regarding valence.
However, all models exhibit poor performance in dominance predictions, with the highest correlation (Qwen-VL-7B) reaching only $\textbf{0.32}$. Most models yield negatively correlated predictions, likely due to their inability to distinguish clearly between expressed and evoked emotions.  
For instance, a character with a fierce facial expression and horrible appearance may \textbf{express} a sense of power and dominance but simultaneously \textbf{evoke} feelings of oppression and low dominance in the viewer, causing contradictory outcomes when analyzed by MLLMs predominantly from the perspective of expressed emotions, rather than the supposed standpoint.
These observations align closely with the findings from the perception task, further highlighting the limitations of current MLLMs in fine-grained emotional analysis and identifying directions for future improvement in understanding evoked emotional attributes.
(We further address failure cases with analysis and improvement suggestions in Supp. \ref{sec: failure cases}.)

\section{Conclusion}

In this paper, we introduce \textbf{EEmo-Bench}, the first benchmark for the comprehensive evaluation of the image-evoked emotions perception and understanding capabilities of MLLMs. 
Our benchmark adopts a carefully curated collection of images from diverse content categories, with rigorous manual annotations, incorporating emotion ranking and the VAD model to assess the diversity and intensity of evoked emotions.
Then, a four-dimensional evaluation framework is proposed, including \textit{Perception}, \textit{Ranking}, \textit{Description}, and \textit{Assessment}. 
Experimental results from evaluating $19$ MLLMs demonstrate significant variability in model performance across different tasks, with a consistent deficiency in accurately perceiving comprehensive image-evoked emotions and understanding specific emotional attributes.
The \textbf{EEmo-Bench} offers critical insights and suggests promising directions for future advancements in the MLLMs IEA field, particularly in enhancing their ability to achieve a thorough and nuanced understanding of image-evoked emotions.


\begin{acks}
This work was supported in part by the National Natural Science Foundation of China under Grant 62271312 and Grant 62132006, and in part by STCSM under Grant 22DZ2229005. 
\end{acks}

\bibliographystyle{ACM-Reference-Format}
\balance
\bibliography{sample-base}

\clearpage
\appendix

\section{Supplementary Material} \label{supp}

\subsection{Details of Annotation Process} \label{supp: annotation}

\begin{figure*}[htb]
\centering
\begin{subfigure}{0.58\textwidth}
    \centering
    \includegraphics[height=4cm, keepaspectratio]{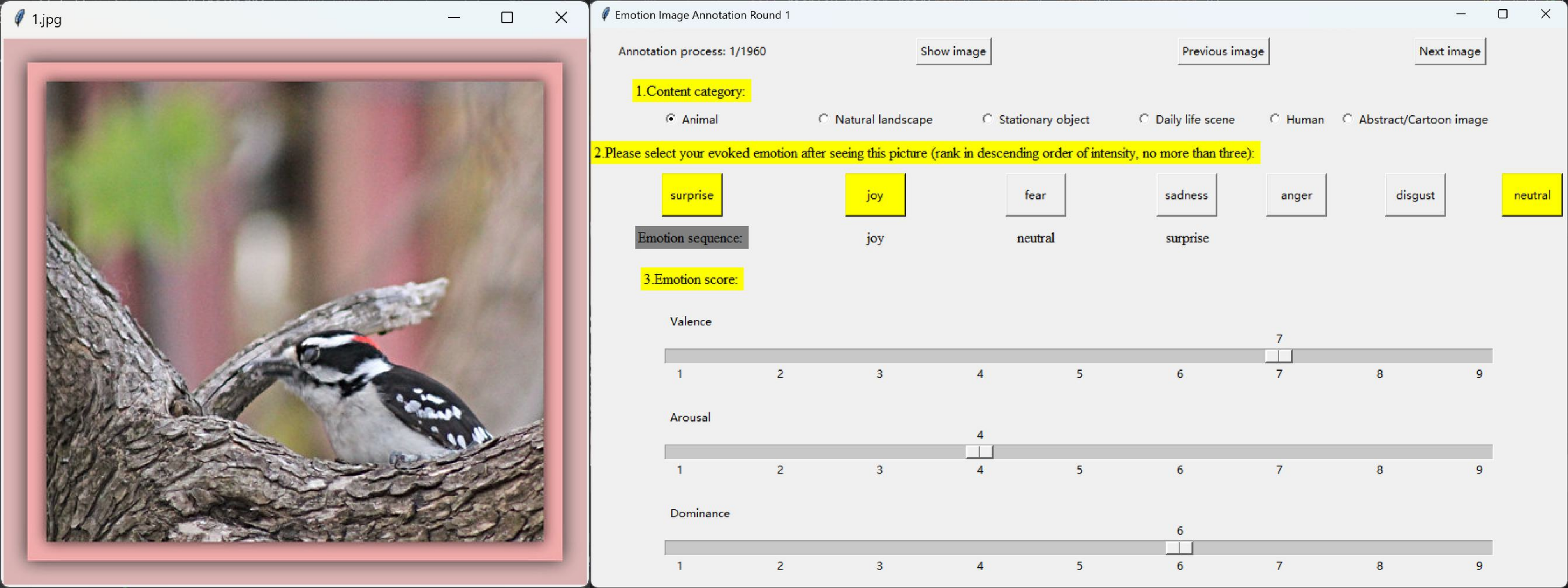}
    \caption{Annotation GUI for emotion attributes of each image}
    \label{fig:GUI 1}
\end{subfigure}
\hfill
\begin{subfigure}{0.4\textwidth}
    \centering
    \includegraphics[height=4cm, keepaspectratio]{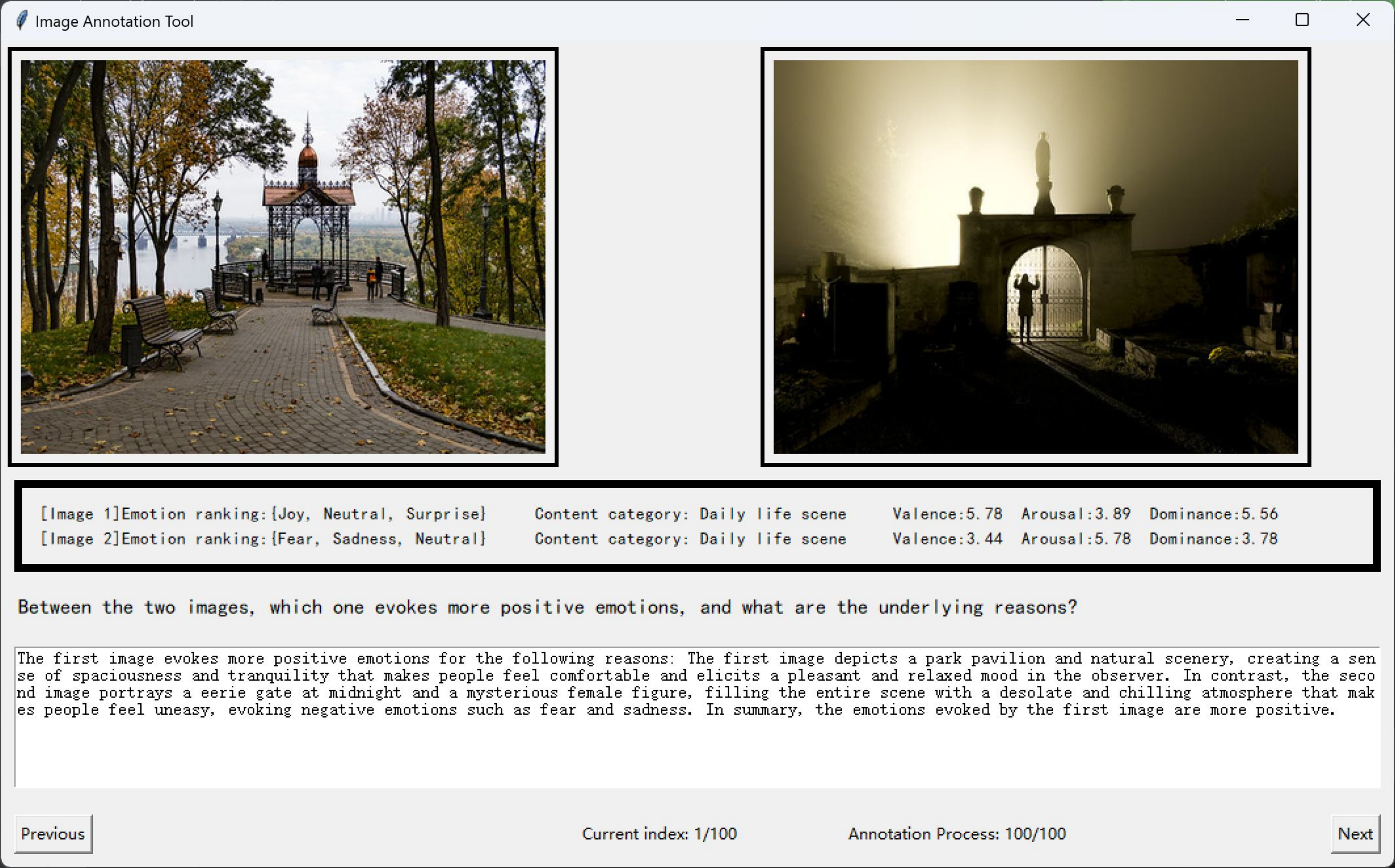}
    \caption{Annotation GUI for description task of \textbf{EEmo-Bench}}
    \label{fig:GUI 2}
\end{subfigure}
\caption{Illustration of the annotation GUI for \textbf{EEmo-Bench}. 
 (a) is used for labeling emotional attributes, including the main evoked emotions (up to three, ranked by intensity) and corresponding VAD scores.
 (b) shows the GUI for collecting golden descriptions, with image pairs open-ended question annotation as an example.}
\end{figure*}

The annotation process takes place in a well-controlled laboratory environment to guarantee silence and consistency. Before starting each experiment, we provide a ten-minute break for participants to enter a stable and calm mental state, which is vital for emotional analysis. To ensure the reliability and quality of the annotation, the entire experiment process is divided into $6$ phases, with no more than one and a half hours per phase and no less than half an hour between each phase. The annotation interface is shown in Fig. \ref{fig:GUI 1}
For the assessment of Valence-Arousal-Dominance (VAD), we enumerate some synonyms and a detailed description of each level from the score $1$ to $9$, with five statements of different levels for each attribute.
The descriptions are elaborated as follows:

 (1) \textbf{valence} measures the positivity or negativity of emotions, reflecting the degree of pleasure experienced by the viewer in a given context.

\begin{itemize}
    \item \textbf{1}: Represents extremely negative emotions or intense displeasure (e.g., sadness, anger, disgust, etc.).
    \item \textbf{3}: Represents relatively strong negative emotions or moderate displeasure (e.g., fear, discomfort, surprise, etc.).
    \item \textbf{5}: Represents neutral emotions, which are neither positive nor negative (e.g., neutral affect).
    \item \textbf{7}: Represents relatively positive emotions or moderate pleasure (e.g., surprise, contentment, etc.).
    \item \textbf{9}: Represents extremely positive emotions or intense pleasure (e.g., joy, satisfaction, etc.).
\end{itemize}

 (2) \textbf{Arousal}  reflects the intensity of emotional activation, indicating the degree of intense or calmness experienced by the viewer in a given circumstance.

\begin{itemize}
    \item \textbf{1}: Represents extremely low arousal, where the viewer feels very calm, relaxed, or even drowsy (e.g., neutral emotions, contentment).
    \item \textbf{3}: Represents relatively low arousal, where the viewer feels relatively calm and relaxed, with mild emotional reactions (e.g., sadness, pleasure).
    \item \textbf{5}: Represents moderate arousal, which is neither intense nor calm (e.g., neutral emotions).
    \item \textbf{7}: Represents relatively high arousal, where the viewer feels excited, nervous, or strongly affected (e.g., happiness, disgust, anger).
    \item \textbf{9}: Represents extremely high arousal, where the viewer feels very excited, tense, energetic, or intensely emotional (e.g., surprise, fear, delight, anger).
\end{itemize}

 (3) \textbf{Dominance} refers to the perceived sense of control or mastery experienced by the individual within a given context, indicating whether they feel dominant or subordinate in the situation.

\begin{itemize}
    \item \textbf{1}: Extremely low dominance – The individual feels entirely dominated by the environment or others, experiencing emotions such as fear and helplessness.
    \item \textbf{3}: Relatively low dominance – The individual feels significantly influenced by the environment or others, with emotions like fear, sadness, disgust, or surprise.
    \item \textbf{5}: Moderate dominance – The individual feels neither fully controlled nor entirely in control, experiencing neutral emotions or a balanced state.
    \item \textbf{7}: Relatively high dominance – The individual feels a strong sense of control over their emotions, with emotions that are more actively initiated, such as satisfaction, pleasure, or anger.
    \item \textbf{9}: Extremely high dominance – The individual feels complete control over everything, associated with emotions such as confidence and excitement.
\end{itemize}

For the description task in \textbf{EEmo-Bench}, we additionally recruited four participants to annotate golden descriptions, with the GUI interface shown in Fig. \ref{fig:GUI 2}.
All annotators primarily relied on the emotional attribute labels obtained from the preliminary annotations to answer open-ended questions. This approach minimizes individual subjective emotional biases, thereby simulating the emotional analysis patterns of the majority of people.
To ensure the accuracy and rigor of the golden description, each participant is supposed to annotate no more than 30 images a day to avoid fatigue, and every result is meticulously checked and revised by another participant before being adopted.

To ensure inter-annotator agreement (IAA), we have trained annotators with expertise. The degree of agreement can be quantitatively assessed using the Intraclass Correlation Coefficient (ICC), which is calculated as follows:

\begin{equation}
    ICC=\frac{MS_B-MS_W}{MS_B+(k-1)MS_W+\frac{k}{n}(MS_R-MS_W)},
\end{equation}

\noindent where $k$ represents the number of raters, $n$ is the sample size, $MS_B$ denotes the between-group variance, $MS_W$ is the within-group variance, and $MS_R$ refers to the rater variance. ICC values for valence, arousal, and dominance are $0.9885$, $0.9414$, and $0.9710$, respectively, indicating strong consistency between annotators.

\subsection{Benchmark Setting}

\subsubsection{\textbf{Prompt for Single Images on Perception Task}} \hfill

\textit{\#User: Assume you are an expert in emotional psychology. Analyze the following emotional psychology questions based on the image. The involved emotion categories include anger, disgust, fear, joy, neutral, sadness, and surprise. The question is as follows: [Question] Choose between one of the following options: [Options]}

\subsubsection{\textbf{Prompt for Image Pairs on Perception Task}} \hfill \label{supp: perception_pair_prompt}

\textit{\#User: Assume you are an expert in emotional psychology. Based on the two given images, please analyze the following emotional psychology questions. Be sure to understand the emotional attributes each image evokes in the viewer and the emotional relationship between these two images in your answers. The emotions to be considered include anger, disgust, fear, joy, neutral, sadness, and surprise. The questions are as follows: [Question] Choose between one of the following options: [Options]}

\subsubsection{\textbf{Prompt for Emotion Ranking Task}} \hfill

\textit{\#User: Assume you are an expert in emotional psychology. Based on the given image, carefully identify the three main emotions evoked by the image, ranking them from the strongest to the weakest. The emotions to consider are limited to: anger, disgust, fear, joy, neutral, sadness, and surprise. Show all three main emotions and use commas to separate them in [].}

\subsubsection{\textbf{Prompt for Single Images on Description Task}} \hfill

\textit{\#User: Assume you are an expert in emotional psychology. Based on the given image, analyze the given image and answer the following question or request regarding emotional responses. The emotions to consider include: anger, disgust, fear, joy, neutral, sadness, and surprise. Please provide insightful and structured responses based on your analysis. The question or request is shown in []: [Question]}

\subsubsection{\textbf{Prompt for Image Pairs on Description Task}} \hfill

\textit{\#User: Assume you are an expert in emotional psychology. Based on the two given images, analyze the images and answer the following question or request regarding emotional responses. Be sure to understand the three main emotions evoked by each image and the emotional relationship between both images. The emotions to consider include: anger, disgust, fear, joy, neutral, sadness, and surprise. Please provide insightful and structured responses based on your analysis. The question or request is shown in []: [Question]}

\begin{table*}[htbp]
    \label{code: vad}
  \begin{tabular}{lp{10cm}}
    \toprule
    \textbf{Algorithm 1} Pytorch-style Pseudo Code on Assessment Task with MLLMs \\
    \midrule
    \begin{lstlisting}[language=Python, basicstyle=\ttfamily\footnotesize, keywordstyle=\color{blue}, stringstyle=\color{red}]
from MLLM_model import Model, Tokenizer, embed_image_and_text
from PIL import Image
import torch

vad_dimension = "valence" # arousal, dominance
prompt = f"Assume you are an expert in emotional psychology. How would you rate the {vad} this " \
"image evokes in the viewer? The level of {vad} this image evokes in the viewer is"

model, tokenizer = Model, Tokenizer()
image = Image.open("my_image.jpg")
input_embeds = embed_image_and_text(image, prompt)
output_logits = model(input_embeds=input_embeds)["logits"][:, -1]
toks = ["Positive", "Neutral", "Negative"] # related level keywords
ids_ = [id_[0] for id_ in tokenizer(toks)["input_ids"]]
p_l = torch.softmax(output_logits[:, ids_], -1))
weight = torch.tensor([1,0.5,0])
vad_pred = torch.inner(p_l, weight)
    \end{lstlisting} \\
    \bottomrule
  \end{tabular}
  
\end{table*}

\subsubsection{\textbf{Example Code on Assessment Task}} \hfill




In Algo. $1$, we provide an example code for evaluating the assessment abilities of MLLMs for different attributes.
It should be noted that the pseudo code can be easily integrated with any new MLLMs (based on the transformers architecture), enabling them to output quantified predictions of the VAD value.

\subsection{Evaluation Details} \label{supp: evaluation}

\subsubsection{\textbf{Evaluation on Perception Task}} \hfill \label{supp: eval_perception}

For the perception task in \textbf{EEmo-Bench}, accuracy is adopted to represent the performance of MLLMs.
While most MLLMs provide a direct option letter in their response, some express their choice through a detailed analysis, which complicates the direct extraction of answers. 
In the EEmo-Bench, the Deepseek-assisted extraction strategy (using Deepseek-V3) is utilized to identify the option most aligned with the MLLMs' reasoning from the extended analysis. 
Additionally, we conduct five rounds of evaluation to minimize random errors. 
For samples that yield consistent judgments across the five rounds, we retain the results, while any discrepancies in judgments are manually reviewed, significantly improving both efficiency and accuracy.
The prompt for judging is shown as follows (taking the prompt for the single image as an example):

\textit{\#User: You will now be provided with a question [question] and a set of options [answers] with option [correct\_ans] being the correct answer. Additionally, there will be an answer [answer] provided by a respondent. Please determine whether the respondent's answer is correct, considering the context of the question. Even if the word choice is not completely the same, you can decide based on the given options
and see whether the one in the answer is close enough to the given correct answer. The result is 1 if the answer
is correct, and otherwise the result is 0. It should be noted that if the answer does not express a clear attitude or 
relevant keywords, or if it means that it is impossible to give a single and clear option judgment,
the result will be 0. Please only provide the result in the following format: Score:}

\subsubsection{\textbf{Perception Task evaluation example}} \hfill

\textit{You will now be provided with a question [Which part of the image evokes the feeling of surprise that you perceive?] and a set of options
    [A: Unique shooting angle, B: Characters' facial expressions, C: The movements of the person.] with option [B: Characters' facial expressions] being the correct answer. Additionally, there will be an answer [The image predominantly evokes the feeling of surprise through the characters' facial expressions. ]
    provided by a respondent. Please determine whether the respondent's answer is correct, considering the context
    of the question. Even if the word choice is not completely the same, you can decide based on the given options
    and see whether the one in the answer is close enough to the given correct answer. The result is 1 if the answer
    is correct, and otherwise the result is 0. It should be noted that if the answer does not express a clear attitude or 
    relevant keywords, or if it means that it is impossible to give a single and clear option judgment,
    the result will be 0. Please only provide the result in the following format: Score:}

\textit{\textbf{5-round Deepseek Score: [1, 1, 1, 1, 1]}}

\subsubsection{\textbf{Evaluation on Ranking Task}} \hfill

Some MLLMs fail to produce outputs in the specified format for a given prompt, for instance, generating emotion categories outside the predefined set of emotions in \textbf{EEmo-Bench}, or responding with long descriptions of emotional reactions, making it challenging to directly extract valid outputs from the model response.
To address this, we also employ Deepseek-assisted extraction strategy, with the prompt in the following format:

\textit{\#User: You will now be provided with a request [question] and a response [emotion\_str] provided by 
a respondent. Please modify the content in the response to the form required by the request, and pay 
attention to the following matters: 1. The emotions you extract must be within the 7 emotions considered 
in the request(anger, disgust, fear, joy, neutral, sadness, and surprise), otherwise they cannot be output. 
If it is a different part of speech form of one of the 
emotions, it can be converted and retained; 2. Extract emotions according to the meaning of the response, 
without mixing in your subjective feelings; 3. If more than three emotions are listed in the response, 
only up to 3 emotions can be displayed in order of intensity; 4. If there are fewer than three emotions 
according to the evaluation criteria, only those that meet the criteria should be displayed, and there 
should be no duplication in the display; 
5. If there is no emotion that meets the criteria, it will be 
returned as an empty list []. Please only provide the result in the following format: Emotion list:}

After obtaining the final predicted emotion list, we apply the ranking strategy outlined in Sec. \ref{sec: ranking} to compute the emotion ranking score for each image.

\subsubsection{\textbf{Ranking Task Evaluation Example}} \hfill

\textit{\#User: You will now be provided with a request [Assume you are an expert in emotional psychology. Based on the given image, carefully identify the three main emotions evoked by the image, ranking them from the strongest to the weakest. The emotions to consider are limited to: anger, disgust, fear, joy, neutral, sadness, and surprise. Show all three main emotions and use commas to separate them in [].] and a response [[Neutral, surprise, awe]] provided by 
a respondent. Please modify the content in the response to the form required by the request, and pay 
attention to the following matters: 1. The emotions you extract must be within the 7 emotions considered 
in the request(anger, disgust, fear, joy, neutral, sadness, and surprise), otherwise they cannot be output. 
If it is a different part of speech form of one of the 
emotions, it can be converted and retained; 2. Extract emotions according to the meaning of the response, 
without mixing in your subjective feelings; 3. If more than three emotions are listed in the response, 
only up to 3 emotions can be displayed in order of intensity; 4. If there are fewer than three emotions 
according to the evaluation criteria, only those that meet the criteria should be displayed, and there 
should be no duplication in the display; 
5. If there is no emotion that meets the criteria, it will be 
returned as an empty list []. Please only provide the result in the following format: Emotion list:}

\textit{\textbf{Final emotion list:[Neutral, Surprise]}}

\subsubsection{\textbf{Evaluation on Description Task}} \label{supp: eval_description} \hfill

We also employ the Deepseek-assisted methodology for evaluating open-ended questions. For efficiency, each question-answer (Q\&A) pair is scored by DeepSeek-V3 based on three dimensions: \textbf{Completeness}, \textbf{Accuracy}, and \textbf{Relevance}, with scores in the set {0, 1, 2}. Each sample is evaluated five times to enhance discriminability and reduce evaluation bias. Let $S_k$ represent the score assigned by Deepseek-V3 in the $k$-th round, where $S_k \in {0, 1, 2}$ and $k = 1, 2, 3, 4, 5$. We then calculate the average of these five scores, $S_k$, and normalize the result using the following formula to obtain $S_{ave}$:

\begin{equation*}
    S_{ave}=\frac{1}{2}\cdot\frac{1}{5}\sum_{k=1}^5 S_k.
\end{equation*}

This normalization maps the final score to a range between $0$ and $1$. We adopt $S_{ave}$ of each sample for final accuracy representation on the description task.
The prompt for evaluation on open-ended questions displays as follows:

\textit{\#User: Given the question [question], evaluate whether the response [answer] completely matches the correct 
answer [correct\_ans]. It should be noted that the emotions considered during the evaluation include 
anger, disgust, fear, joy, neutral, sadness, and surprise. The redundancy or absence of other emotional words is 
not used as a criterion for judging the accuracy of the answer.
First, check the response, and please rate the score 0 if the response is not a valid answer.
Please rate score 2 if the response completely or almost completely matches the correct answer on completeness, accuracy, and relevance. 
Please rate score 1 if the response partly matches the correct answer on completeness, accuracy, and relevance.
Please rate score 0 if the response doesn't match the correct answer on completeness, accuracy, and relevance at all.
Please only provide the result in the following format: Score:}


\subsubsection{\textbf{Description Task Evaluation Example}} \hfill

\textit{\#User: Given the question [Do you think the emotional transmission in this image is direct or indirect? Please explain your view.], evaluate whether the response [The emotional transmission in this image appears to be indirect. The gorilla depicted is not displaying any overt facial expressions that would directly convey emotions such as anger, disgust, fear, joy, sadness, or surprise. Instead, the emotional response is inferred through the context and the viewer's interpretation of the gorilla's posture, environment, and the overall atmosphere of the scene.] completely matches the correct 
answer [The emotional conveyance of this image is quite direct. The gorilla, positioned at the center of the frame, exudes a sense of unease and fear through its majestic posture and piercing gaze directly into the lens. Its angry expression is effectively communicated to the viewer, showcasing a strong sense of presence and power. The forest background and the gorilla's natural state also directly convey a sense of wildness and primitive beauty. The overall composition is clear, with the subject prominently highlighted, and the emotional expression is unambiguous.]. It should be noted that the emotions considered during the evaluation include 
anger, disgust, fear, joy, neutral, sadness, and surprise. The redundancy or absence of other emotional words is 
not used as a criterion for judging the accuracy of the answer.
First, check the response, and please rate the score 0 if the response is not a valid answer.
Please rate score 2 if the response completely or almost completely matches the correct answer on completeness, accuracy, and relevance. 
Please rate score 1 if the response partly matches the correct answer on completeness, accuracy, and relevance.
Please rate score 0 if the response doesn't match the correct answer on completeness, accuracy, and relevance at all.
Please only provide the result in the following format: Score:}

\textit{\textbf{5-round Deepseek score: [0,1,0,0,0]}}
\textit{\textbf{Final description score: $\textbf{1/10=0.1}$}}

\begin{figure}
  \centering
  \includegraphics[width=\linewidth]{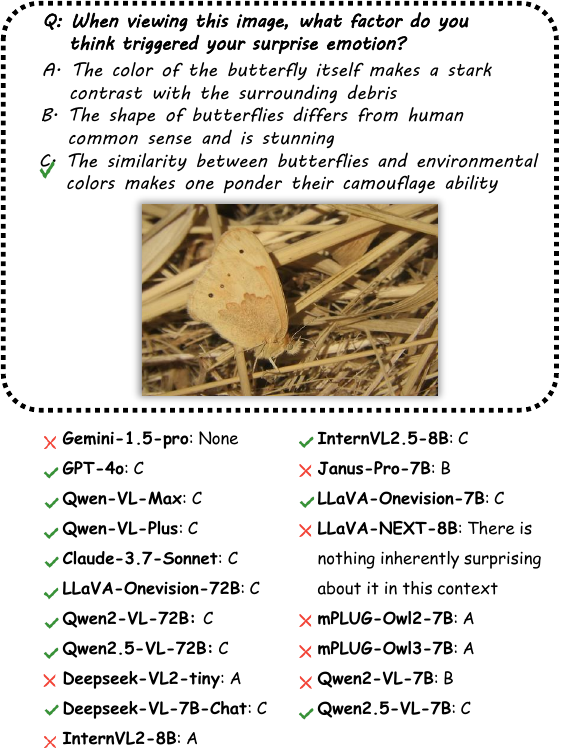}
  \caption{Qualitative comparison of MLLMs on perception task.}
  \label{fig: perception_example}
\end{figure}

\subsection{Visualization Examples of Results}

\begin{figure*}
  \includegraphics[width=0.7\textwidth]{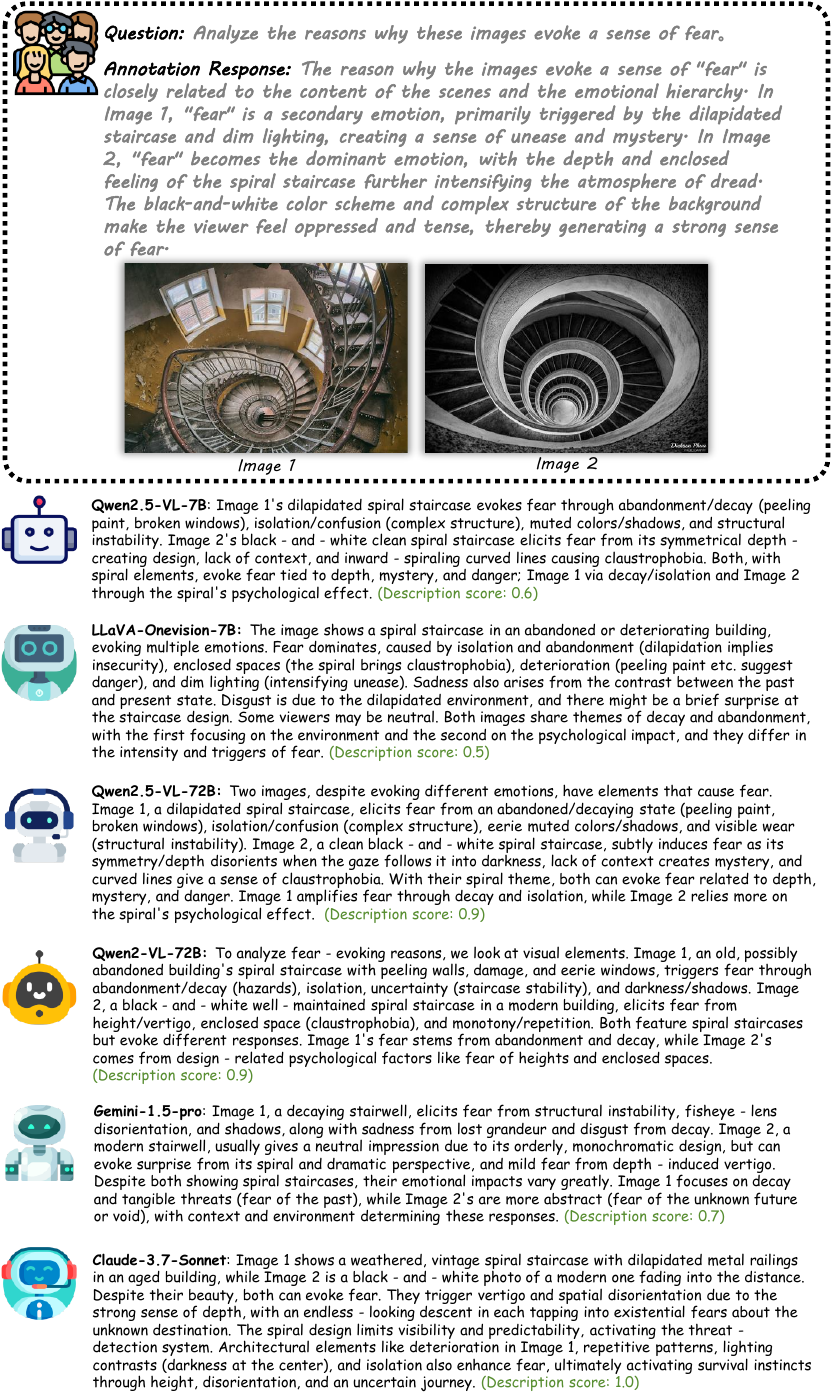}
  \caption{A qualitative comparison of MLLMs on the description task, with the description score for each model provided following the response.
  }
  \Description{examples}
  \label{fig:description_examples}
\end{figure*}

\begin{figure*}
  \includegraphics[width=0.8\textwidth]{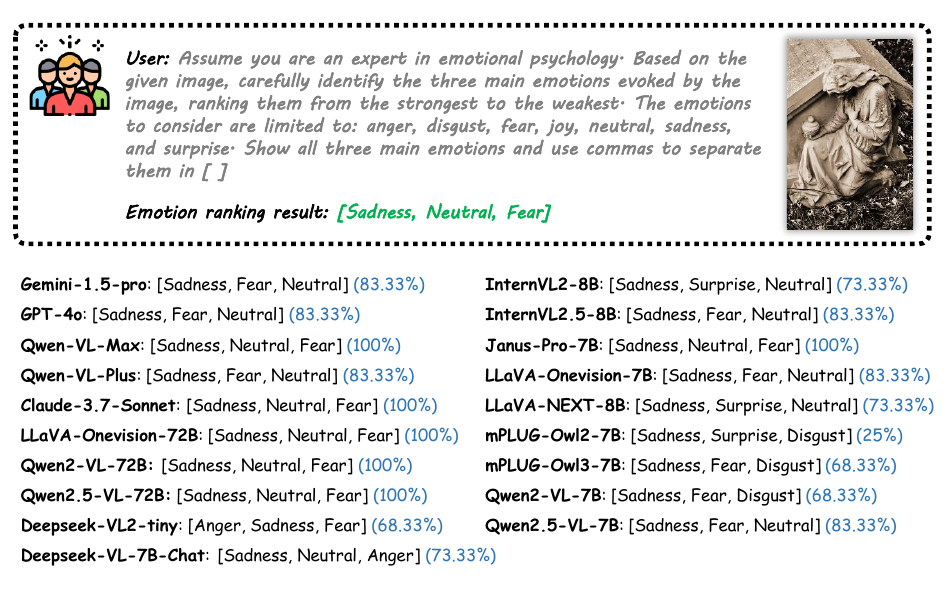}
  \caption{A qualitative comparison of MLLMs on the ranking task, with the emotion ranking score for each model provided following the response.
  }
  \Description{examples}
  \label{fig:ranking_examples}
\end{figure*}

In Fig. \ref{fig: perception_example}, we present an example of MLLMs' responses to the perception task on single images. Using a Deepseek-assisted strategy to extract the correct answer letter from the model output (\textit{see Supp. \ref{supp: evaluation} for details}), most MLLMs provide valid responses. However, LLaVA-NEXT-8B and Gemini-pro-1.5 seem to lack reasoning ability in this example.
The results highlight that for reasoning perception of evoked emotions, a task relatively simple for humans, only about half of the MLLMs provide the correct answer. Notably, most of the models that perform well are large-scale open-source or proprietary MLLMs, underscoring the need for improvements in MLLMs' emotional perception capabilities.

In Fig. \ref{fig:description_examples}, we present example outputs for open-ended questions on image pairs from the top 2 performing models in each MLLM category, accompanied by the final scores evaluated by Deepseek-V3. (\textit{See Supp. \ref{supp: eval_description} for details about the evaluation approach}.)
The results reveal significant differences in the open-ended question answering ability of current MLLMs for emotion analysis. In this example, proprietary models, particularly Claude-3.7-Sonnet, demonstrate the closest resemblance to human-like reasoning in understanding the causes of emotions. 
However, for medium-scale MLLMs such as LLaVA-Onevision-7B, their ability to perform emotional reasoning still requires further improvement. 
The performance disparities in open-ended question answering highlight the current instability in MLLMs' understanding of evoked emotions.

In Fig. \ref{fig:ranking_examples}, we present an example that demonstrates the ability of different MLLMs to rank the emotions evoked by images. The results show that all MLLMs can detect one to two primary evoked emotions. 
Among them, both proprietary and large-scale open-source MLLMs exhibit the highest average ranking performance, with many of these models accurately predicting the emotional ranking in this example, highlighting their promising capability in emotion perception. 
However, giving the majority of MLLMs can recognize the dominant emotion of an image, many models reverse the order of primary emotions or provide unrelated emotions, indicating significant limitations in their ability to fully understand the emotions of an image.

\begin{figure}
  \centering
  \includegraphics[width=\linewidth]{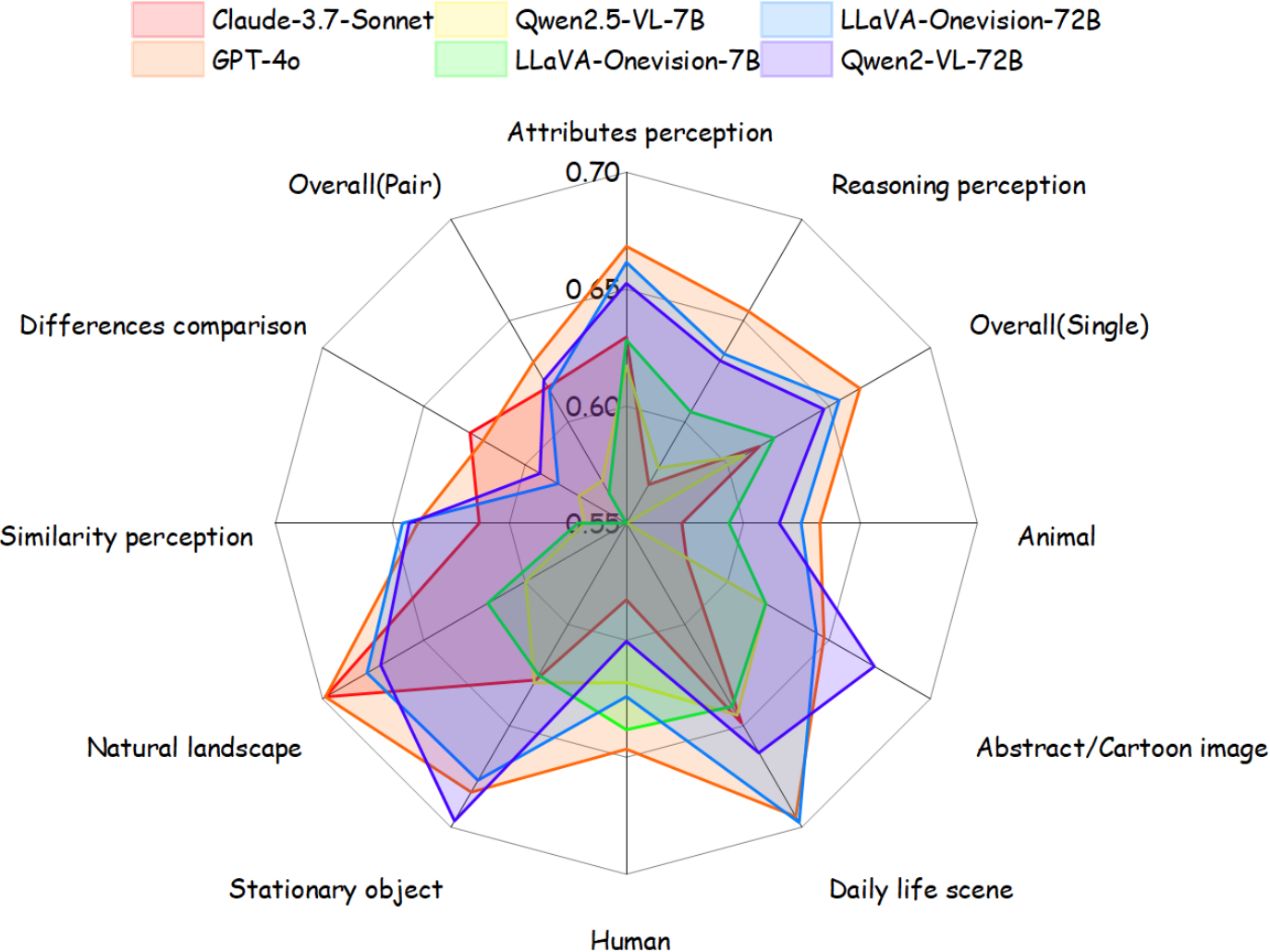}
  \caption{Radar chart comparing the results of the two-dimensional concerns and six content categories in the single-image emotion perception task, as well as the two-dimensional concerns in the image-pair emotion perception task for MLLMs.}
  \label{fig: perception_example}
\end{figure}

\subsection{Extended Results for Perception} \label{supp: extended results}

Supplementary experiments are conducted to evaluate two independent perspectives defined in the perception task (see Sec. \ref{sec: perception}): the categorization by question concerns and by image content category. The performance across these different dimensions can be quickly visualized in Fig. \ref{fig: perception_example}.
The findings are illustrated as follows:

\textbf{(1) Question concerns in single images}.
From Tab. \ref{tab: extended}, we observe that MLLMs exhibited similar performance in both \textit{attributes perception} and \textit{reasoning perception} for single images (except for Gemini-1.5-pro and Claude-3.7-Sonnet), indicating comparable abilities in perceiving emotions and reasoning the causes of emotional responses. 
Among them, GPT-4o achieved the best performance on both question concerns, with scores of $66.84\%$ and $65.45\%$, respectively, demonstrating its balanced understanding capabilities in single-image emotion analysis.

\textbf{(2) Question concerns in image pairs}.
As shown in Table \ref{tab: extended}, MLLMs’ performance varies significantly across different question concerns in image pairs. Large-scale open-source MLLMs excel in similarity perception, demonstrating an advantage in recognizing coarse-grained emotional polarity and conducting joint analyses. 
In contrast, proprietary MLLMs, particularly Gemini-1.5-pro and Claude-3.7-Sonnet, outperform in comparing emotional differences in image pairs. Despite the higher task difficulty, their superior performance highlights their strength in fine-grained emotional perception, such as comparing arousal intensity.

\textbf{(3) Performance across image content categories}.
The performance results in Table \ref{tab: extended} show significant variation in MLLMs' ability to perceive emotions across different image content categories. 
Specifically, images depicting daily life scenes and natural landscapes generally yield higher accuracy in emotion perception tasks, suggesting that emotions and their attributes in these images are more easily captured by MLLMs. This may stem from MLLMs' stronger emotional understanding of specific events and typical elements, such as scene colors. 
Notably, large-scale open-source MLLMs perform better at emotion perception for stationary objects, indicating a greater ability to understand symbolic meanings. 
However, most MLLMs exhibit a notable gap in understanding the emotions evoked by animal images, highlighting a weaker ability to incorporate human common sense and preferences. This suggests a direction for future research on the emotional perception capabilities of MLLMs across different image categories.

\begin{table*}[]
\footnotesize
\setlength{\tabcolsep}{2pt} 
\caption{Result of the two-dimensional concern and six content categories on the single images emotions perception task, and the two-dimensional concern on image pairs emotions perception task of MLLMs. The best performance is \textbf{bolded} and the second and third performances are {\ul underlined.}}
\begin{tabular}{l|ccc|c@{\hspace{1pt}}c@{\hspace{1pt}}cccc|ccc}
\hline
\multicolumn{1}{c|}{\textbf{Sub-categories}}                          & \multicolumn{3}{c|}{\textbf{Single Image}}                                                                                                                                          & \multicolumn{6}{c|}{\textbf{Content category}}                                                                                                                                                                                                                                                                                      & \multicolumn{3}{c}{\textbf{Image Pair}}                                                                                                                           \\ \hline

\multicolumn{1}{c|}{{\scriptsize \textbf{MLLMs}}}                                   & \textit{\begin{tabular}[c]{@{}c@{}}{\scriptsize Attributes}\\ {\scriptsize perception↑}\end{tabular}} & \textit{\begin{tabular}[c]{@{}c@{}}{\scriptsize Reasoning}\\ {\scriptsize perception↑}\end{tabular}} & \multicolumn{1}{c|}{\textit{{\scriptsize Overall↑}}} & \textit{{\scriptsize Animal↑}} & \textit{\begin{tabular}[c]{@{}c@{}}{\scriptsize Abstract/}\\ {\scriptsize Cartoon image↑}\end{tabular}} & \textit{\begin{tabular}[c]{@{}c@{}}{\scriptsize Daily} \\ {\scriptsize life scene↑}\end{tabular}} & \textit{{\scriptsize Human↑}}  & \textit{\begin{tabular}[c]{@{}c@{}}{\scriptsize Stationary}\\ {\scriptsize object↑}\end{tabular}} & \multicolumn{1}{c|}{\textit{\begin{tabular}[c]{@{}c@{}}{ \scriptsize Natural}\\ {\scriptsize landscape↑}\end{tabular}}} & \textit{\begin{tabular}[c]{@{}c@{}}{\scriptsize Similarity}\\ {\scriptsize perception↑}\end{tabular}} & \textit{\begin{tabular}[c]{@{}c@{}}{\scriptsize Differences} \\ {\scriptsize comparison↑}\end{tabular}} & \textit{{\scriptsize Overall↑}} \\ \hline
\multicolumn{1}{c|}{\textbf{Random guess}}                            & 42.22\%                                                          & 39.83\%                                                         & \multicolumn{1}{c|}{41.67\%}          & 41.67\%          & 41.67\%                                                            & 41.67\%                                                      & 41.67\%          & 41.67\%                                                      & \multicolumn{1}{c|}{41.67\%}                                                      & 46.39\%                                                          & 36.81\%                                                            & 42.13\%          \\ \hline
\multicolumn{13}{l}{\textit{\scriptsize Medium-scale open-source MLLMs}}                                                                                                                                                                                                                                                                                                                                                                                                                                                                                                                                                                                                                                                                      \\ \hline
\multicolumn{1}{l|}{\scriptsize Deepseek-VL2-tiny}                       & 50.98\%                                                          & 50.22\%                                                         & \multicolumn{1}{c|}{50.81\%}          & 48.12\%          & 51.53\%                                                            & 53.58\%                                                      & 48.94\%          & 51.92\%                                                      & \multicolumn{1}{c|}{51.96\%}                                                      & 45.92\%                                                          & 41.63\%                                                            & 44.41\%          \\
\multicolumn{1}{l|}{\scriptsize Deepseek-vl-chat-7B}                     & 55.92\%                                                          & 53.20\%                                                         & \multicolumn{1}{c|}{55.16\%}          & 52.82\%          & 54.98\%                                                            & 57.45\%                                                      & 55.08\%          & 52.31\%                                                      & \multicolumn{1}{c|}{59.11\%}                                                      & 50.61\%                                                          & 39.85\%                                                            & 45.80\%          \\
\multicolumn{1}{l|}{\scriptsize InternVL2-8B}                            & 55.59\%                                                          & 56.62\%                                                         & \multicolumn{1}{c|}{55.83\%}          & 51.74\%          & 57.09\%                                                            & 57.31\%                                                      & 57.68\%          & 56.35\%                                                      & \multicolumn{1}{c|}{55.03\%}                                                      & 54.92\%                                                          & 50.83\%                                                            & 53.26\%          \\
\multicolumn{1}{l|}{\scriptsize InternVL2.5-8B}                          & 57.40\%                                                          & 56.62\%                                                         & \multicolumn{1}{c|}{57.22\%}          & 51.61\%          & 56.90\%                                                            & 60.74\%                                                      & 56.97\%          & 56.54\%                                                      & \multicolumn{1}{c|}{61.39\%}                                                      & 58.58\%                                                          & 56.32\%                                                            & 57.84\%          \\
\multicolumn{1}{l|}{\scriptsize Janus-Pro-7B}                            & 52.54\%                                                          & 55.30\%                                                         & \multicolumn{1}{c|}{53.18\%}          & 53.08\%          & 50.96\%                                                            & 55.30\%                                                      & 50.24\%          & 55.00\%                                                      & \multicolumn{1}{c|}{55.37\%}                                                      & 49.86\%                                                          & 45.47\%                                                            & 48.33\%          \\
\multicolumn{1}{l|}{\scriptsize LLaVA-Onevision-7B}                      & 62.83\%                                                          & 60.49\%                                                         & \multicolumn{1}{c|}{62.29\%}          & 59.38\%          & 61.88\%                                                            & 64.04\%                                                      & {\ul 63.83\%}    & 62.50\%                                                      & \multicolumn{1}{c|}{61.84\%}                                                      & 57.08\%                                                          & 54.92\%                                                            & 56.49\%          \\
\multicolumn{1}{l|}{\scriptsize LLaVA-NEXT-8B}                           & 55.26\%                                                          & 61.26\%                                                         & \multicolumn{1}{c|}{56.65\%}          & 55.90\%          & 52.87\%                                                            & 60.60\%                                                      & 49.17\%          & 59.04\%                                                      & \multicolumn{1}{c|}{64.91\%}                                                      & 50.98\%                                                          & 45.34\%                                                            & 49.48\%          \\
\multicolumn{1}{l|}{\scriptsize mPLUG-Owl2-7B}                           & 55.86\%                                                          & 54.97\%                                                         & \multicolumn{1}{c|}{55.66\%}          & 54.29\%          & 51.92\%                                                            & 58.31\%                                                      & 54.14\%          & 54.81\%                                                      & \multicolumn{1}{c|}{60.48\%}                                                      & 55.48\%                                                          & 41.38\%                                                            & 49.52\%          \\
\multicolumn{1}{l|}{\scriptsize mPLUG-Owl3-7B}                           & 55.59\%                                                          & 57.51\%                                                         & \multicolumn{1}{c|}{56.04\%}          & 51.88\%          & 52.68\%                                                            & 59.17\%                                                      & 54.61\%          & 58.65\%                                                      & \multicolumn{1}{c|}{60.31\%}                                                      & 55.76\%                                                          & 57.09\%                                                            & 56.34\%          \\
\multicolumn{1}{l|}{\scriptsize Qwen2-VL-7B}                             & 59.51\%                                                          & 60.26\%                                                         & \multicolumn{1}{c|}{59.69\%}          & 57.77\%          & 60.34\%                                                            & 60.89\%                                                      & 59.46\%          & 60.00\%                                                      & \multicolumn{1}{c|}{60.14\%}                                                      & 52.67\%                                                          & 56.58\%                                                            & 54.70\%          \\
\multicolumn{1}{l|}{\scriptsize Qwen2.5-VL-7B}                           & 61.73\%                                                          & 57.73\%                                                         & \multicolumn{1}{c|}{60.81\%}          & 54.69\%          & 61.88\%                                                            & 64.47\%                                                      & 61.82\%          & 62.88\%                                                      & \multicolumn{1}{c|}{59.97\%}                                                      & 56.79\%                                                          & 57.34\%                                                            & 57.09\%          \\ \hline
\multicolumn{1}{c}{\textit{\scriptsize Large-scale   open-source MLLMs}} &                                                                  &                                                                 &                                       &                  &                                                                    &                                                              &                  &                                                              &                                                                                   &                                                                  &                                                                    &                  \\ \hline
\multicolumn{1}{l|}{\scriptsize LLaVA-Onevision-72B}                     & {\ul 66.15\%}                                                    & 63.36\%                                                         & \multicolumn{1}{c|}{{\ul 65.50\%}}    & {\ul 62.47\%}    & {\ul 64.37\%}                                                      & \textbf{69.77\%}                                             & 62.41\%          & 67.69\%                                                      & \multicolumn{1}{c|}{67.80\%}                                                      & \textbf{64.57\%}                                                 & 58.37\%                                                            & 61.56\%          \\
\multicolumn{1}{l|}{\scriptsize Qwen2-VL-72B}                            & {\ul 65.25\%}                                                    & 63.02\%                                                         & \multicolumn{1}{c|}{{\ul 64.74\%}}    & {\ul 61.53\%}    & \textbf{67.24\%}                                                   & 66.33\%                                                      & 60.05\%          & \textbf{69.69\%}                                             & \multicolumn{1}{c|}{67.12\%}                                                      & {\ul 64.29\%}                                                    & 59.26\%                                                            & {\ul 62.06\%}    \\
\multicolumn{1}{l|}{\scriptsize Qwen2.5-VL-72B}                          & 64.09\%                                                          & {\ul 64.46\%}                                                   & \multicolumn{1}{c|}{64.18\%}          & 59.38\%          & 63.60\%                                                            & 66.76\%                                                      & 61.23\%          & {\ul 69.23\%}                                                & \multicolumn{1}{c|}{67.46\%}                                                      & {\ul 64.29\%}                                                    & 60.03\%                                                            & {\ul 61.86\%}    \\ \hline
\multicolumn{1}{l}{\textit{\scriptsize Proprietary   MLLMs}}             &                                                                  &                                                                 &                                       &                  &                                                                    &                                                              &                  &                                                              &                                                                                   &                                                                  &                                                                    &                  \\ \hline
\multicolumn{1}{l|}{\scriptsize Gemini-1.5-pro}                          & 64.42\%                                                          & 59.82\%                                                         & \multicolumn{1}{c|}{63.36\%}          & 56.17\%          & 63.41\%                                                            & {\ul 68.39\%}                                                & {\ul 62.65\%}    & 62.50\%                                                      & \multicolumn{1}{c|}{{\ul 68.26\%}}                                                & 58.82\%                                                          & {\ul 60.31\%}                                                      & 59.64\%          \\
\multicolumn{1}{l|}{\scriptsize GPT-4o}                                  & \textbf{66.84\%}                                                 & \textbf{65.45\%}                                                & \multicolumn{1}{c|}{\textbf{66.53\%}} & \textbf{63.27\%} & {\ul 64.75\%}                                                      & {\ul 69.48\%}                                                & \textbf{64.66\%} & {\ul 68.27\%}                                                & \multicolumn{1}{c|}{\textbf{69.85\%}}                                             & 63.92\%                                                          & {\ul 62.07\%}                                                      & \textbf{62.95\%} \\
\multicolumn{1}{l|}{\scriptsize Qwen-VL-Max}                                & 63.49\%                                                          & {\ul 64.56\%}                                                   & \multicolumn{1}{c|}{63.74\%}          & 59.65\%          & 64.04\%                                                            & 65.57\%                                                      & 61.10\%          & 68.09\%                                                      & \multicolumn{1}{c|}{66.44\%}                                                      & 63.29\%                                                          & 59.31\%                                                            & 61.22\%          \\
\multicolumn{1}{l|}{\scriptsize Qwen-VL-Plus}                               & 59.87\%                                                          & 58.80\%                                                         & \multicolumn{1}{c|}{59.63\%}          & 55.23\%          & 58.08\%                                                            & 63.61\%                                                      & 59.31\%          & 62.45\%                                                      & \multicolumn{1}{c|}{59.80\%}                                                      & 59.60\%                                                          & 57.38\%                                                            & 59.06\%          \\
\multicolumn{1}{l|}{\scriptsize Claude-3.7-Sonnet}                       & 62.96\%                                                          & 56.91\%                                                         & \multicolumn{1}{c|}{61.56\%}          & 57.37\%          & 57.97\%                                                            & 64.85\%                                                      & 58.27\%          & 62.74\%                                                      & \multicolumn{1}{c|}{\textbf{69.85\%}}                                             & 61.29\%                                                          & \textbf{62.71\%}                                                   & 61.71\%          \\ \hline
\end{tabular}
\label{tab: extended}
\end{table*}

\begin{figure*}[t]
  \centering

  \begin{subfigure}[t]{\textwidth}
    \centering
    \includegraphics[width=\textwidth]{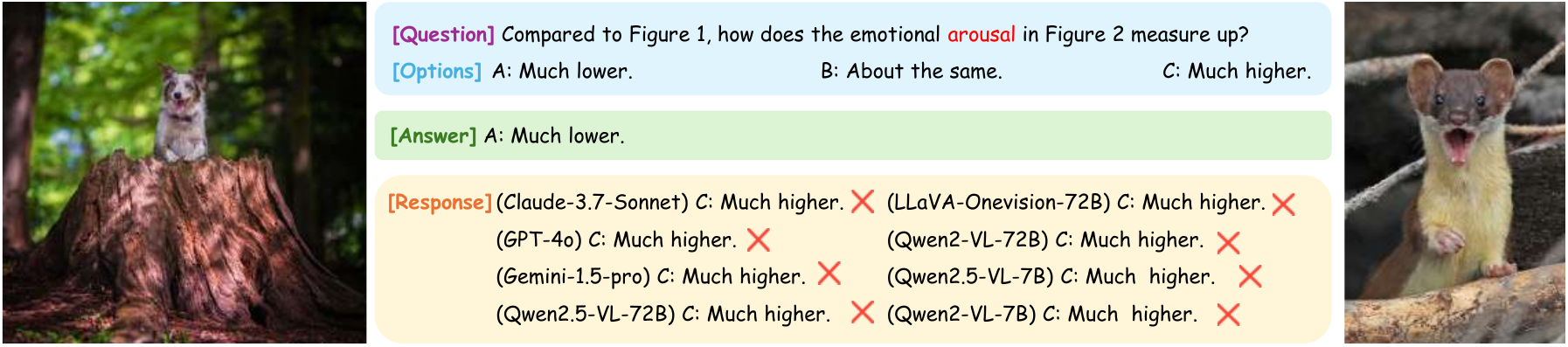}
    \caption{Arousal analysis.}
    \label{fig:fc1}
  \end{subfigure}

  \vspace{4pt}

  \begin{subfigure}[t]{\textwidth}
    \centering
    \includegraphics[width=\textwidth]{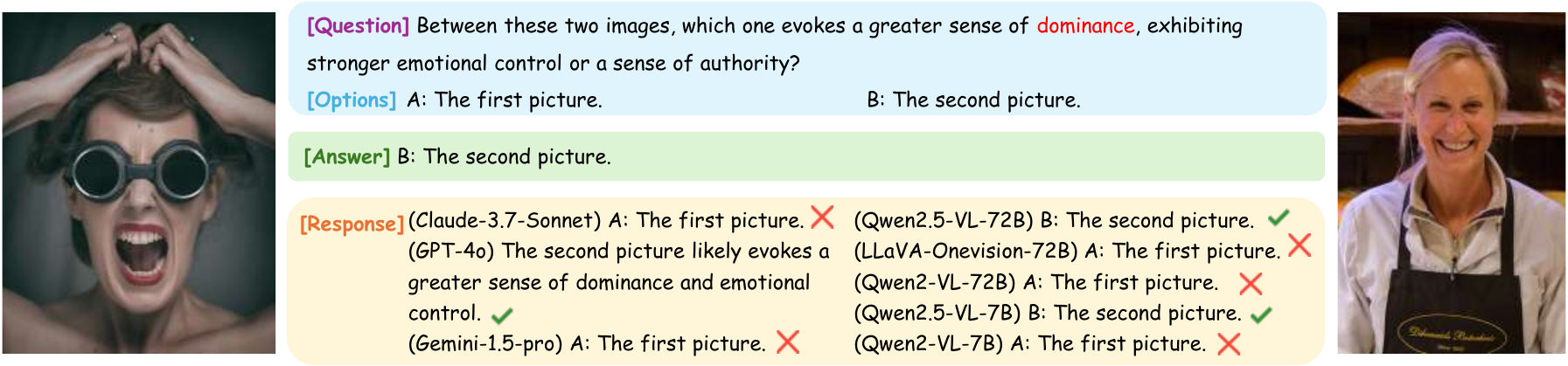}
    \caption{Dominance analysis.}
    \label{fig:fc2}
  \end{subfigure}

  \vspace{4pt}

  \begin{subfigure}[t]{\textwidth}
    \centering
    \includegraphics[width=\textwidth]{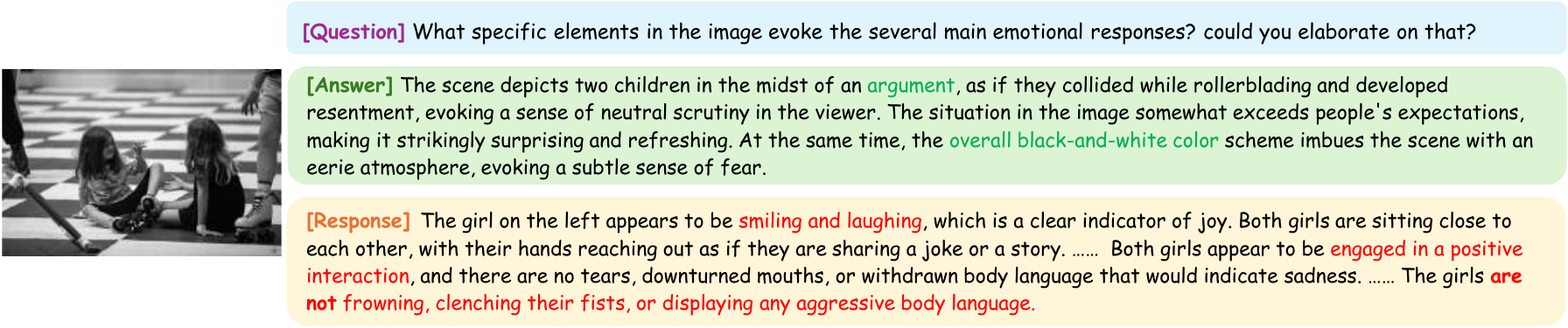}
    \caption{Perceptual focus misalignment}
    \label{fig:fc3}
  \end{subfigure}

  \vspace{4pt}

  \begin{subfigure}[t]{\textwidth}
    \centering
    \includegraphics[width=\textwidth]{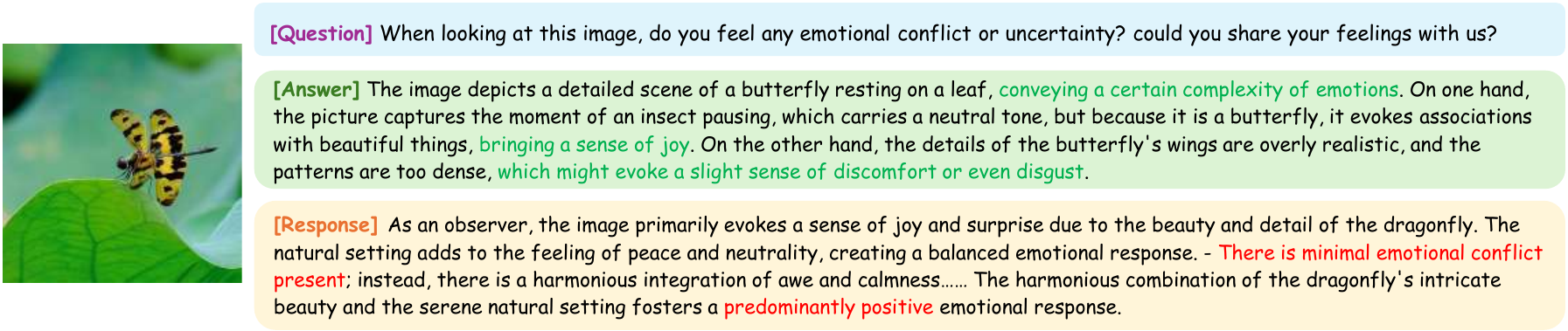}
    \caption{Insufficient Common Sense Understanding}
    \label{fig:fc4}
  \end{subfigure}

  \caption{Typical failure cases of MLLMs on perception and description tasks.}
  \label{fig:failure cases}
\end{figure*}

\subsection{Observation on Failure Cases}
\label{sec: failure cases}

\subsubsection{\textbf{Arousal Analysis}} \hfill

To visually illustrate the difference in MLLMs' emotional understanding between single images and image pairs, as discussed in Sec. \ref{sec: observation}, we present an analysis of a representative failure case.
We evaluate the token limits for all open-source models, ruling out context-window issues and attributing the results to models' inherent limitations. 
We find that MLLMs' understanding of emotional intensity and diversity remains coarse, often recognizing only primary emotions and VAD polarity, preventing nuanced comparisons. 
As shown in Fig.~\ref{fig:fc1}, in the comparison of arousal, the left image—featuring a cute puppy with harmonious colors and a shallow depth of field—presents a rich and novel visual experience that stimulates viewer interest. In contrast, the right image, which depicts a surprised mole with an open mouth but relatively muted tones, typically evokes lower arousal level. 
However, nearly all models provide incorrect predictions, suggesting that MLLMs tend to focus more on the emotional expression of the main subject when interpreting arousal, while remaining less sensitive to visual cues such as color and lighting that can elicit high-arousal emotions like curiosity and surprise.

\subsubsection{\textbf{Dominance Analysis}} \hfill

To support our argument on the performance of dominance assessment discussed in Sec.\ref{sec: observation}, we further analyze a failure case illustrated in Fig.\ref{fig:fc2}. 
In the left image, a woman shouting conveys a strong expression of dominance but simultaneously induces a sense of oppression in the viewer. In contrast, the right image, featuring a warm smile, expresses moderate dominance yet evokes a heightened sense of psychological dominance, such as warmth and satisfaction. Nevertheless, all models incorrectly classify the left image as eliciting higher dominance.

\subsubsection{\textbf{Improvement Suggestions}} \hfill

Regarding the discussion on enhancing the emotion understanding capabilities of MLLMs in images, we provide a reasoned analysis of the differences in quantitative scores and accuracy in Sec. \ref{sec: observation}. 
Here, we focus on the discussion of failure cases, providing additional detailed analysis and improvement suggestions as follows. 

\textbf{(1) Three-step perception.}
After analyzing some failure cases, we find that a three-step COT reasoning approach could help MLLMs better understand the evoked emotions. The process involves: \textbf{first, identifying typical emotional features} (e.g., characters' expressions); \textbf{second, perceiving the overall event}; and \textbf{third, incorporating low-level features} like color for judgment. 
For instance, in Fig. \ref{fig:fc3}, two children are sitting on an ice rink, waving their arms, with one girl displaying expressions of anger and pain. Despite this, most MLLMs interpret the scene as playful, associating it with positive emotions such as "joy." However, a more nuanced interpretation suggests that the children may have collided and blamed each other, which would likely evoke "anger" in the girl. This example highlights the importance of deeper contextual understanding in the emotional interpretation of scenes.

\textbf{(2) Commonsense reasoning.}
MLLMs need to improve their understanding of human common sense and preferences, particularly when it comes to animals. Incorporating social media comments during the pre-training stage of MLLMs could aid in this process. As shown in Supp. \ref{supp: extended results} of the paper, even advanced models like GPT-4o and Claude-3.7-Sonnet struggle with interpreting emotions related to animals. For instance, as illustrated in Fig. \ref{fig:fc4}, while a close-up of a dragonfly with natural coloring is often linked to positive emotions by MLLMs, it may evoke discomfort for many people, thereby reducing the image's overall valence. This underscores the importance of understanding people's inherent impressions and emotional responses to various stimuli, an aspect that MLLMs must learn to interpret more accurately.

\subsection{Limitation}

\noindent \textbf{Subjectivity in annotation}: 
Although we recruit participants with diverse educational backgrounds, genders, and ages, and employ multiple rounds of annotation and filtering, individual subjective preferences and cultural differences may still influence overall emotional responses. 
This could lead to the amplification of certain emotions or biases in understanding specific emotional attributes, a challenge inherent in the field of emotion analysis.

\noindent \textbf{Number of Emotions in Ranking}: The ranking strategy employed in the benchmark aims to capture emotional diversity by retaining the top three emotions ranked by intensity. 
However, not every image generates three emotions with significant intensity differences. For instance, in images where the dominant evoked emotion is `neutral', the lower-ranked emotions may have weak intensities and minimal differences, which limits the effectiveness of the fixed number of emotions based on weighted scores as described in Sec. \ref{sec: ranking}. 
In future work, we intend to explore a methodology to dynamically determine the number of emotions involved in the ranking based on their weighted scores, thus maximizing emotional diversity while reducing perceptual ambiguity.  

\end{document}